\documentclass[aps,prd,showpacs,showkeys,superscriptaddress,twocolumn]{revtex4-1}
\usepackage[colorlinks,linkcolor=blue,anchorcolor=blue,citecolor=blue,urlcolor=blue,breaklinks=true]{hyperref}
\usepackage{graphicx}
\usepackage{amsmath}
\usepackage{amssymb}
\usepackage{slashed}
\usepackage{latexsym}
\usepackage{epsfig}
\usepackage{amsbsy}
\usepackage{array}
\usepackage{amssymb}
\usepackage{setspace}
\usepackage{bm}
\usepackage{lipsum}
\usepackage{mathrsfs}
\usepackage{float}
\usepackage{color}
\usepackage{booktabs}
\usepackage[T1]{fontenc}
\usepackage{mathptmx}
\usepackage{changes}
\DeclareMathAlphabet{\mathcal}{OMS}{cmsy}{m}{n}
\DeclareSymbolFont{largesymbols}{OMX}{cmex}{m}{n}

\begin{document}
\author{Li-Qun Su}
\email{xzslq1203@smail.nju.edu.cn}
\affiliation{Department of physics, Nanjing University, Nanjing 210093, China}
\author{Chao Shi}
\email{cshi@nuaa.edu.cn}
\affiliation{Department of Nuclear Science and Technology, Nanjing University of Aeronautics and Astronautics, Nanjing 210016, China}
\author{Yong-Feng Huang}
\email{hyf@nju.edu.cn}
\affiliation{School of Astronomy and space science, Nanjing University, Nanjing 210023, China}
\author{Yan Yan}
\email{2919ywhhxh@163.com}
\affiliation{School of Mathematics and physics, Changzhou University,Changzhou 213164, China}
\author{Cheng-Ming Li}
\email{licm.phys@gmail.com}
\affiliation{School of physics and Microelectronics, Zhengzhou University, Zhengzhou 450001, China}

\author{Hongshi Zong}
\email{zonghs@nju.edu.cn}
\affiliation{Department of physics, Nanjing University, Nanjing 210093, China}
\affiliation{Department of physics, Anhui Normal University, Wuhu 241000, China}
\affiliation{Nanjing Proton Source Research and Design Center, Nanjing 210093, China}
\affiliation{Joint Center for Particle, Nuclear Physics and Cosmology, Nanjing 210093, China}
\date{\today}
	
\title{Hybrid stars can be self-bound}
	
\begin{abstract}
Based on the properties of uniform nuclear matter at the nuclear saturation density and basic thermodynamic relations, we first re-study the composition of matter on the surface of normal neutron stars and hybrid stars. It is found that hybrid stars are composed of uniform hadronic matter on the surface rather than heavy nuclei. Then we use the Walecka model and the self-consistent NJL model to describe the equation of state of low-density hadrons and quark matter at high densities respectively. The P--interpolation method is employed to connect the equation of state at the extreme densities to study hybrid stars. As a result, we find that the obtained hybrid star mass-radius relation and tidal deformability meet the requirements of the latest astronomical data. More importantly, we find that the hybrid stars we obtained can be self-bound rather than gravitationally bound, which is completely different from previous related studies.
\bigskip

\end{abstract}
	
\maketitle
\section{Introduction}

Neutron stars are some of the densest manifestations of massive objects in the universe.
Remarkable progresses have been made in neutron-star physics, covering from many directions \cite{glendenning2012compact,PhysRevC.52.2072,Baym_2018,Serot:1984ey,PhysRevC.78.015802,PhysRevC.88.025801,Chamel:2008ca,PhysRevC.94.035804,PhysRevC.97.035805,PhysRevLett.119.161101,nature} and providing new observational windows into the microscope physics of dense strongly interacting matter. First, as the discovery of massive neutron stars, i.e., PSR J0348 + 0432 \cite{Antoniadis1233232} with 2.01$\pm$0.04 $M_{\odot}$, many models with a soft equation of state (EOS) are excluded because they can not support such massive stars. Second, x-ray observations provided much more precise measurements of the radius \cite{Bogdanov_2019,Riley_2019,Capano_2020}, such as $11.0^{+0.9}_{-0.6}$~km for the 1.4-solar-mass neutron stars in Ref. \cite{Capano_2020}. Third, due to the gravitational wave detection, the binary neutron star (BNS) merger event GW170817 \cite{PhysRevLett.119.161101} supplies constraints on the tidal deformability, $\Lambda<800$ for the 1.4-solar-mass neutron stars, which rules out models with much stiff EOS. These astronomical observations give us new ways of inferring both equilibrium and dynamical properties of neutron stars.

Although people have accumulated a lot of astronomical observational data about neutron stars, people still have great controversy about the internal structure of neutron stars.
 At present, it is generally believed that there are three types of neutron star structure models.
The first is to assume that the neutron star is composed of hadronic matter, and there is no asymptotically free quark matter inside. The second is to assume that when the core density of a traditional neutron star is large enough, a hybrid star/ mixed star with a core of quark matter will be formed \cite{PhysRevD.98.083013,PhysRevD.97.103013,PhysRevD.92.054012}. The third is the so-called ``strange quark matter'' star. Following the argument from Witten \cite{PhysRevD.30.272} that strange quark matter is more stable, the existence of quark star is discussed by many authors \cite{PhysRevC.89.055207, PhysRevD.101.063023,PhysRevD.43.627,PhysRevC.62.025801,PhysRevD.92.084009,PhysRevD.99.043001}.  In recent years, many authors have proposed the possible existence of two-flavor quark stars \cite{PhysRevD.100.043018,PhysRevD.100.123003}. As discussed in Ref.  \cite{PhysRevLett.120.222001}, two-flavor quark matter can be the stable ground state for the baryon number $A > 300$ after taking the bulk effect into account. In this study, we mainly discuss the so-called hybrid star, that is, the outer region of the compact star is composed of uniform hadronic matter, and the inner core is composed of asymptotically free non-strange quark matter.

It is generally believed that the surface pressure and baryon density of ordinary neutron stars and hybrid stars are zero, and the region of $0.3\sim0.5~\mathrm{n}_0$ is made of inhomogeneous hadronic matter in a ``pasta'' phase \cite{PhysRevC.88.025801, Chamel:2008ca, PhysRevC.94.035804, PhysRevC.99.025804, PhysRevC.97.035805}, where $\mathrm{n}_0$ is the nuclear saturation density. As the baryon number density goes beyond $0.5~\mathrm{n}_0$, the nuclei begin to coalesce into the uniform neutron matter. The thermodynamic properties of strongly interacting matter can be effectively illustrated in the mass-radius diagram. For instance, the EOS of free particles requires no particles should exist at zero pressure. Hence, the mass-radius diagram under such an EOS shows that as the mass of a compact star decreases \cite{PhysRevD.98.083013,PhysRevD.97.103013,PhysRevD.92.054012}, the radius of the compact star grows, which reflects the fact that a small mass can no longer constrain a cluster of particles and they tend to expand into the whole space.  However, for self-bound particles, the EOS will be quite different, and the extremely large radius of the  mass-radius diagram naturally vanishes. In this case, for a small mass object, although the gravity itself might be too weak to constraint the particles, the self-bound feature ensures that all particles will be restricted inside the boundary. It explains why quark stars have different mass-radius relation as compared with neutron stars and hybrid stars \cite{PhysRevD.92.054012,PhysRevD.101.063023}.

The main motivation of this paper is to try to use the known properties of the strongly interacting matter at the nuclear saturation density $\mathrm{n}_0$ and some basic thermodynamic relationships to readdress the properties of the strongly interacting matter at low density and then to study hybrid stars.
According to the fact that the density of nuclear matter saturates at $\mathrm{n}_0$, the energy per baryon $E/A$ at $\mathrm{n}_0$ has a minimum value $(E/A)_\mathrm{min}=\mu_0$ \cite{PhysRevD.58.096007} (see below), which is exactly the baryon chemical potential. It implies that the density converges directly to $\mathrm{n}_0$, where the pressure is zero, while states with the density less than $\mathrm{n}_0$ are unstable. Otherwise, the most stable state will be replaced by such a state. Unlike the past picture of hybrid stars where the pressure vanishes as the density becomes zero, the uniform hadronic matter at large densities is stable to be self-bound now.
However, it seems contradictory that the most stable state of nuclear matter is the iron nucleus, which, with a density lower than  $\mathrm{n}_0$, are not supposed to be stable.
In fact, the statement that iron nucleus is the most stable state is attributed to the fact that the bulk effect ($E_\mathrm{v}=-16~\mathrm{MeV}$), the surface energy ($E_\mathrm{s}=18~A^{-1/3}~\mathrm{MeV}$), and the Coulomb repulsion ($E_\mathrm{c}=0.7~Z^2~A^{-4/3}$, where $Z$ is the number of protons), impose a huge influence on $E/A$ for a finite baryon number $A$. On the contrary, for neutron stars, the baryon number $A$ is large enough so that the bulk effect is insignificant and the strong interaction dominates. Under this situation, the uniform hadronic matter appears only when the  baryon number density is larger than $\mathrm{n}_0$.

Due to the fact that uniform hadronic matter is the most stable state of strongly interacting matter, neutron stars or hybrid stars without inhomogeneous hadronic matter on the crust are possible to exist in the universe. With the self-bound hadronic matter on the surface, a small mass  compact star will have a limited radius, and it is different from the mass-radius relation of classic neutron stars and hybrid stars taht have increasing radius with decreasing mass.

This paper is organized as follows.
In Section \ref{section:2},
the reason that enormous asymmetric hadronic matter is more stable than iron nuclei is presented.
In Section \ref{section:3} and \ref{section:4},
The Walecka model and a self-consistent NJL model are engaged for the description of hadronic matter and quark matter, respectively.
In Section \ref{section:5},
The P-interpolation method is employed to smoothly connect the EOS between hadronic matter and quark matter, and the mass-radius relations are illustrated, which satisfy the astronomical observations for massive neutron stars.
In Section \ref{section:6},
our conclusions are presented.

\section{The most stable state}\label{section:2}	

Since we can not derive the properties of cold dense matter at finite density from the first principles of QCD, the experiments of nuclear matter at nuclear saturation density $n_0$ provide us the most reliable constraints on the strongly interacting matter. First of all, the Energy per particle has a minimum value at $n_0$,
\begin{align}
\dfrac{\partial \varepsilon/n}{\partial n}\bigg|_{n_0}=0,\label{eq:stability}
\end{align}
here $\varepsilon$ is the energy density of cold dense matter, and $n$ is the baryon number density.
Besides, with the thermodynamical relation of
\begin{align}
\varepsilon=-P+\mu n, \label{eq:thermo relation}
\end{align}
and combining Eq. (\ref{eq:stability}) and Eq. (\ref{eq:thermo relation}), we have
\begin{align}
P(n_0)=0.\label{eq:Pequalszero}
\end{align}
Substituting Eq. (\ref{eq:Pequalszero}) into Eq. (\ref{eq:thermo relation}), we get the baryon chemical potential $\mu_0$ at the nuclear saturation density $n_0$
\begin{align}
\mu_0=\Big(\dfrac{\varepsilon}{n}\Big)_{\mathrm{min}}.
\end{align}
As long as the chemical potential $\mu$ is lower than $\mu_0$, the baryon number density should be zero,
\begin{align}
n(\mu)=0, ~~\mathrm{for} ~\mu < \mu_0. \label{eq:PN}
\end{align}
If the baryon number density $n(\mu)$ is non-zero at $\mu$ < $\mu_0$, then because $\varepsilon(\mu)/n(\mu)>\varepsilon(\mu_0)/ n(\mu_0)$ so that
\begin{align}
-&\dfrac{P(\mu)}{n(\mu)}+\mu>\mu_0>\mu.
\end{align}
we would have,
\begin{align}
P(\mu)<0, ~~\mathrm{for}~ n(\mu)>0,~\mu<\mu_0,
\end{align}
which is unstable. Hence, the baryon number density is expected to be zero until the chemical potential $\mu$ shifts to $\mu_0$ where it is the most stable state. Once the chemical potential reaches $\mu_0$, the baryon number density discontinuously jumps up to $n_0$. As a result, we conclude that the baryon number density does not gradually grow from zero, but appears suddenly to $n_0$.

Since the vacuum pressure is usually divergent and could not be directly measured, people usually assume that the vacuum pressure is equal to
zero \cite{Kapusta:2006pm}.  In fact, with the universal zero-temperature finite-density pressure relationship \cite{PhysRevD.98.083013,PhysRevD.97.103013,PhysRevD.92.054012,PhysRevD.78.054001,Zong:2008zzb},
\begin{align}
P(\mu)=P(\mu=0)+\int_0^{\mu} n(\tilde{\mu})d \tilde{\mu},
\end{align}
and considering the experimental data at the nuclear saturation density of the strongly interacting matter $P(\mu_0)=0$ and Eq. (\ref{eq:PN}), the vacuum pressure becomes
\begin{align}
P(\mu=0)=-\int_0^{\mu_0} n(\tilde{\mu})d \tilde{\mu}=0,
\end{align}
which is model-independent as well. It implies that the no-sea approximation \cite{Kapusta:2006pm} is not necessary but is constrained by the nature.

A question is aroused concerning what is $\mu_0$ for the strongly interacting matter. To answer the question, we need to consider the binding energy defined as the energy per baryon \cite{PhysRevD.58.096007}:
\begin{align}
\dfrac{\varepsilon}{n}=M_N-(NM_N-E)/N = M_N-E_0,
\end{align}
where $M_N$ is the mass of nucleon, $E/V=\varepsilon, N/V=n$, and the binding energy $E_0=(NM_N-E)/N$. Because the most stable state has a minimum $\mu_0$, the binding energy $E_0$ in turn becomes the maximum value. Empirically, we know that the most stable state is the iron nucleus with $N=A=56$ and $E_0=8~\mathrm{MeV}$, where $N$ and $A$ are particle number and baryon number respectively. For a nucleus with finite size, the bulk effect plays a crucial role in the binding energy. In the core of the nucleus where particles are isotropic,  the electromagnetic interaction is weak and can be omitted as comparing with the strong interaction. So, the central density of the nuclei is approximately $\mathrm{n}_0$. However, the situation changes on the surface of the nucleus, where the strong interaction is absent from the outside. Hence, the electromagnetic interaction can no longer be ignored, which leads to the bulk effect of hadronic matter with finite size. The stability of nuclei is thus prominently influenced by the bulk effect. The Weizsacker formula tells us that \cite{PhysRevD.58.096007}
\begin{align}
E_0=a_1-a_2\dfrac{1}{A^{1/3}}-a_3Z^2\dfrac{1}{A^{4/3}},\label{eq:coulomb}
\end{align}
where $a_1\approx16~\mathrm{MeV}, a_2\approx18~\mathrm{MeV}, a_3\approx0.7~\mathrm{MeV}$, and
$Z$ is the number of protons. Thus, for the nuclear matter with a finite size, $E_0$ goes up to the maximum value at $A=56$, $E_0\approx8.7~\mathrm{MeV}$ and $\mu=\mu_0\approx 931~\mathrm{MeV}$, so that the iron nucleus becomes the most stable state.
Nevertheless, the radius of a massive neutron star is of the order of 10 km, and the hadronic matter on the surface of neutron stars is insufficient as comparing with the total baryon number $A$. Therefore, the bulk effect is ignorable. Although the real phase structure on the surface remains unknown, the stable condition can provide some constraints. For a hybrid star, the outer layer is composed of hadronic matter and the inner core is composed of quark matter. They must coincide in the middle-density region. As the density goes down, the quark matter in the core gradually transforms into the uniform hadronic matter through the deconfinement, and then the density of uniform hadronic matter stops declining when the pressure is zero, corresponding to the stable density $\mathrm{n}_0$. No more structures are necessary to cover the uniform hadronic matter. 

\section{Walecka model}\label{section:3}	
Following other researchers, we use the Walecka model to describe the hadronic matter  \cite{glendenning2012compact,walecka1986,fetter2003quantum,PhysRevC.90.055203,
1972PhT....25k..54F,Kapusta:2006pm,LI2008113,FUKUSHIMA201399,PhysRevC.99.025804,PhysRevC.97.035805},
\begin{align}
\mathscr{L}&=\bar\psi(i\slashed{\partial}-M_N + g_\sigma\sigma-g_{\omega}\omega\gamma^0-g_{\rho}\rho\tau_3\gamma^0)\psi\nonumber\\
&+\dfrac12 (\partial_\mu\sigma\partial^\mu\sigma-m_\sigma^2\sigma^2)-\dfrac13 bM_N(g_\sigma\sigma)^3-\dfrac14 c(g_\sigma\sigma)^4\nonumber\\
&-\dfrac14 \omega_{\mu\nu}\omega^{\mu\nu}+ \dfrac12 m_\omega^2\omega^2-\dfrac14 \rho_{\mu\nu}\rho^{\mu\nu}+ \dfrac12 m_\rho^2\rho^2\nonumber\\
&+\dfrac12 \Lambda g_{\rho}^2g_{\omega}^2\rho^2\omega^2,\label{eq:lagrangian}
\end{align}
where $g_\sigma$, $g_\omega$, $g_\rho$ are the nucleon coupling constant for $\sigma$,  $\omega$ and $\rho$ mesons,
$b$ and $c$ are the coupling constants for the self-interactions of $\sigma$ mesons. $\omega_{\mu_\nu}$ and $\rho_{\mu_\nu}$ are defined as
\begin{align}
\omega_{\mu\nu}=\partial_\mu\omega_\nu-\partial_\nu\omega_\mu\\
\rho_{\mu\nu}=\partial_\mu\rho_\nu-\partial_\nu\rho_\mu.
\end{align}

The gap equations are obtained by variation of the Lagrangian:
\begin{align}
&M_N^*=M_N-g_\sigma\sigma,\label{eq:mass}\\
&\mu_p^*=\mu_p-g_\omega\omega-\dfrac12 g_\rho\rho,\\
&\mu_n^*=\mu_n-g_\omega\omega+\dfrac12 g_\rho\rho,\\
&m_\sigma^2\sigma+bM_Ng_\sigma^3\sigma^2+ c g_\sigma^4\sigma^3=g_\sigma n_s,\\
&m_\omega^2\omega+\Lambda g_\omega^2g_\rho^2\rho^2\omega=g_\omega (n_n+n_p),\\
&m_\rho^2\rho+\Lambda g_\omega^2g_\rho^2\omega^2\rho=\dfrac12 g_\rho (n_n-n_p),\label{eq:drho}
\end{align}
where
\begin{align}
&n_s=2N_f\int\dfrac{d^3 p}{(2\pi)^3}\dfrac{M_N^*}{\sqrt{p^2+{M_N^*}^2}}(f^- + f ^+),\\
&n=2N_f\int\dfrac{d^3 p}{(2\pi)^3}(f^- - f ^+).
\end{align}
Here $
f^\pm=\dfrac{1}{1+ \mathrm{exp}[\beta(E_p\pm\mu)]}
$
is the fermion distribution, and $E_p=\sqrt{p^2+{M_N^*}^2}$.
The thermodynamical potentials are required to illustrate which phase is more stable at a certain situation. Then we have
\begin{align}
\Omega=&-2N_f\int\dfrac{d^3 p}{(2\pi)^3}\big\{T\mathrm{ln}(1+ \mathrm{exp}[-\beta(E_p+\mu)])\nonumber\\
&+T\mathrm{ln}(1+ \mathrm{exp}[-\beta(E_p-\mu)])\big\}\nonumber\\
&-\dfrac12 m_\rho^2\rho^2-\dfrac12 m_\omega^2\omega^2-\dfrac12 \Lambda g_{\rho}^2g_{\omega}^2\rho^2\omega^2\nonumber\\
&+\dfrac12 m_\sigma^2\sigma^2 +\dfrac13 b M_Nm_\sigma^3\sigma^3+\dfrac14 c m_\sigma^4\sigma^4.
\end{align}
Following Ref. \cite{Kapusta:2006pm}, we take the parameters as $M_N=939$ MeV, $m_\sigma=550$ MeV, $m_\omega=783$ MeV, $m_\rho=770$ MeV, $b=7.950\times10^{-3}$, and $c=6.952\times 10^{-4}$. We further take  $g_\sigma^2/4\pi=6.003$, $g_\omega^2/4\pi=5.948$, $g_\rho=4.583$ and $\Lambda=8.431$ , then the binding energy is obtained as $E_0=16.317 \mathrm{MeV}$.
And other features at the saturation point are shown in Table \ref{table:experiment}

\begin{table} [H]
	\caption{}
	\begin{center}
		\begin{tabular}{c|c|c|c|c}
			\hline
             \hline
			 $$ & $n_0(\mathrm{fm}^{-3})$  &  $K(\mathrm{MeV})$  & $E_{sym}(\mathrm{MeV})$ &$L(\mathrm{MeV})$ \\\hline
             $\mathrm{MFT}$ &$0.155$       &$258.213$     & $30.8098 $    &  $80.3535$ \\\hline
			 $\mathrm{Empirical}$ &$0.16\pm0.01$       &$240\pm20$     & $31.7\pm3.2 $    &  $58.7\pm28.1$ \\\hline \hline
		\end{tabular}
	\end{center}
\label{table:experiment}
\end{table}

\begin{figure}[H]
	\centering
	\includegraphics[width=1\linewidth]{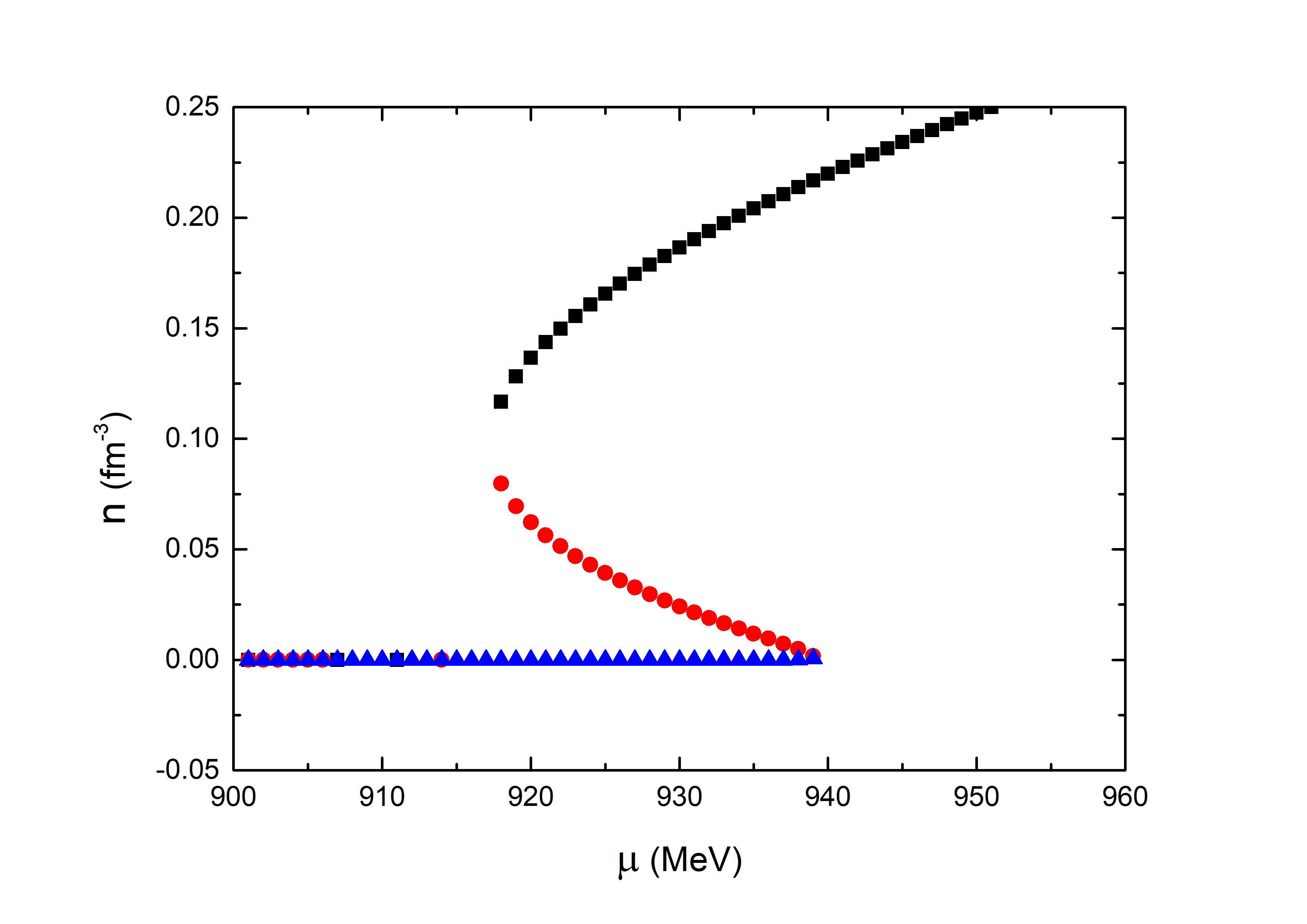}\\
	\caption{The baryon number density as a function of the chemical potential with different initial values.}\label{fig:rho}
\end{figure}

\begin{figure}[H]
	\centering
	\includegraphics[width=1\linewidth]{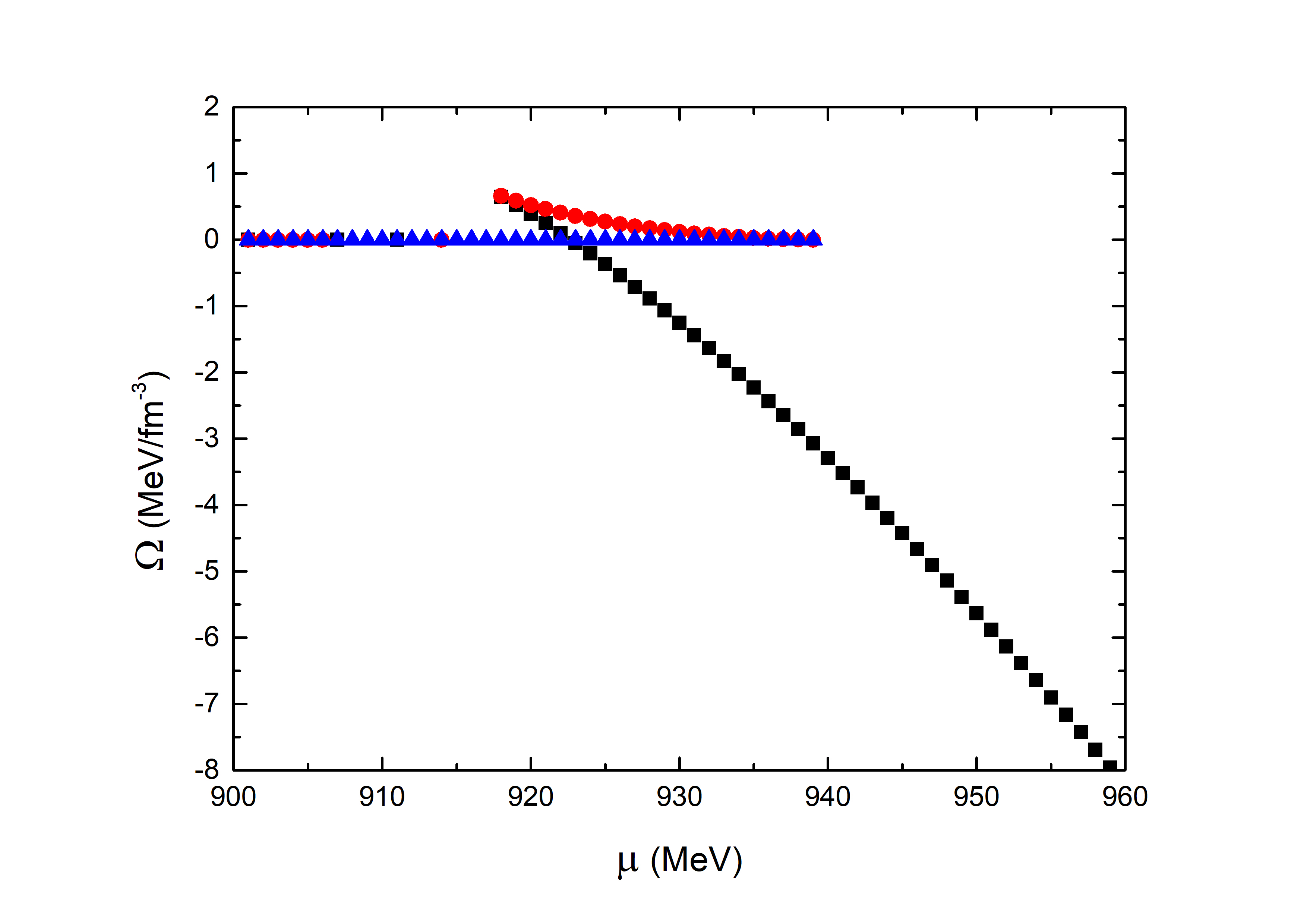}\\
	\caption{The thermodynamical potential as a function of the chemical potential with different initial values.}\label{fig:potential}
\end{figure}

\begin{figure}[H]
	\centering
	\includegraphics[width=1\linewidth]{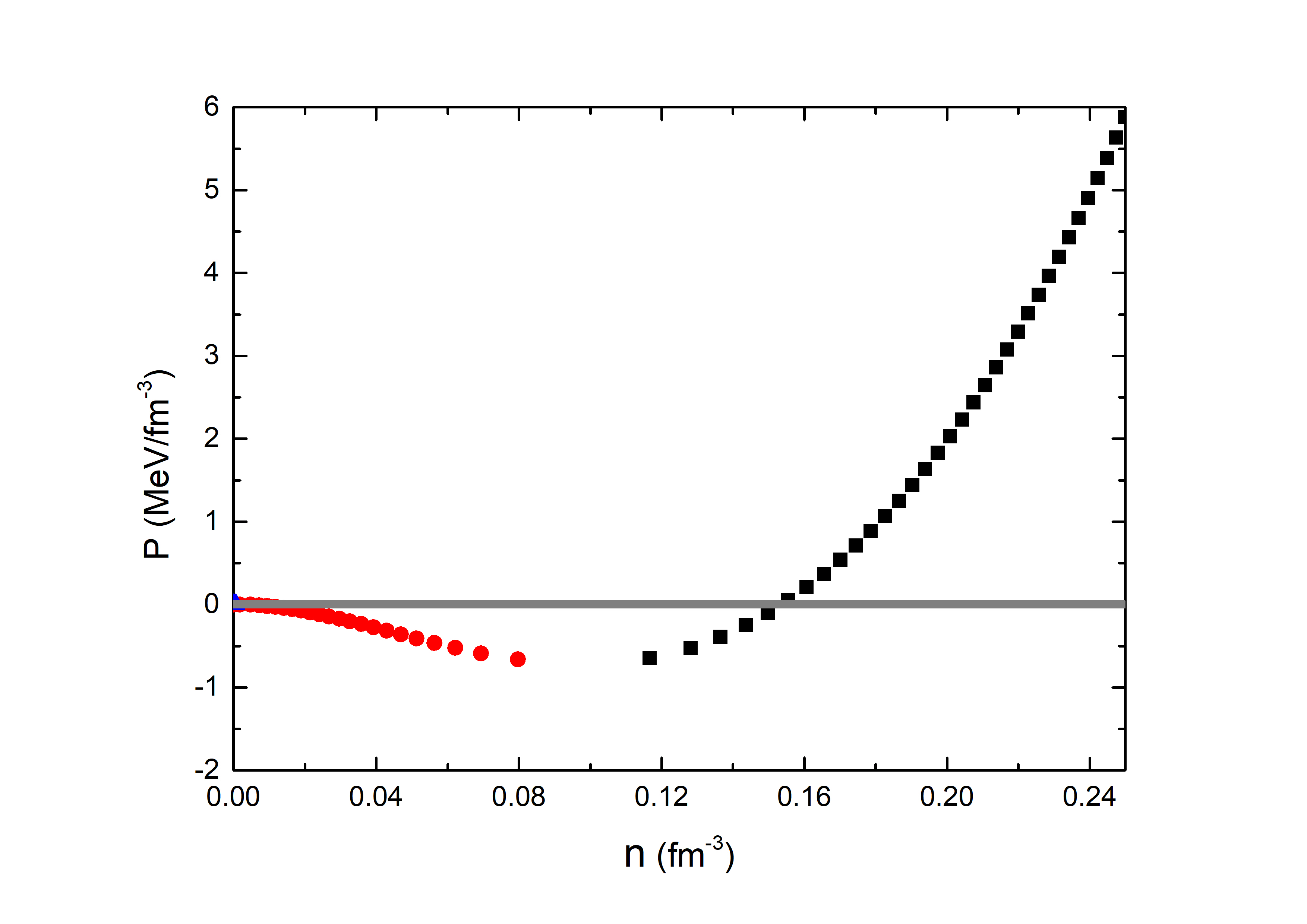}\\
	\caption{The pressure as a function of the baryon number density with different initial values. The grey line corresponds to zero pressure. } \label{fig:pressure}
\end{figure}

The baryon number density is shown in Fig. \ref{fig:rho}. Different colors in the figure represent
the three solutions of gap equations from Eq. (\ref{eq:mass}) - (\ref{eq:drho}) with different
initial values. Unlike the usual picture of free particles that the density increases with the growing
chemical potential, the density of red solutions declines as the chemical potential increases. In fact,
it indicates a liquid-gas phase transition by the Maxwell construction. In order to obtain the physical
EOS of uniform hadronic matter, stable solutions must be distinguished from unstable ones. For this
purpose, the thermodynamical potential of uniform hadronic matter is presented in Fig. \ref{fig:potential}.
It is shown that the red solutions always have a higher thermodynamical potential, and thus it can not be
the most stable states.
As the chemical potential grows, the blue solutions shift to the black one.
The intersection between blue and black solutions implies the position of a first-order phase transition,
$\mu=923~\mathrm{MeV}$, as indicated by the Maxwell construction in Fig. \ref{fig:rho}. Therefore,
in Fig. \ref{fig:rho}, the density of uniform hadronic matter has a singularity at $\mu_0=923~\mathrm{MeV}$.
At low chemical potentials, no particles can be excited from the vacuum, but the density directly jumps
to $\mathrm{n}_0=0.155 ~\mathrm{fm}^{-3}$ as the chemical potential reaches $\mu_0$. Besides, to show
that such solutions are indeed stable, more details about the solutions are shown in
Fig. \ref{fig:pressure}, where the pressure as a function of the baryon number density is plotted.
Apparently, for the red solutions, we have $\partial P/ \partial n<0$. It clearly shows that this is
an unstable state. On the contrary, the black solutions are physical and stable. To sum up the above analysis,
the Walecka model is consistent with the deduction in Section \ref{section:2} that the baryon number density
jumps from zero to the nuclear saturation density $n_0$ when the pressure is zero, and the matter with a
baryon number density less than $n_0$ is unstable and could not exist.

\begin{figure}[H]
	\centering
	\includegraphics[width=1\linewidth]{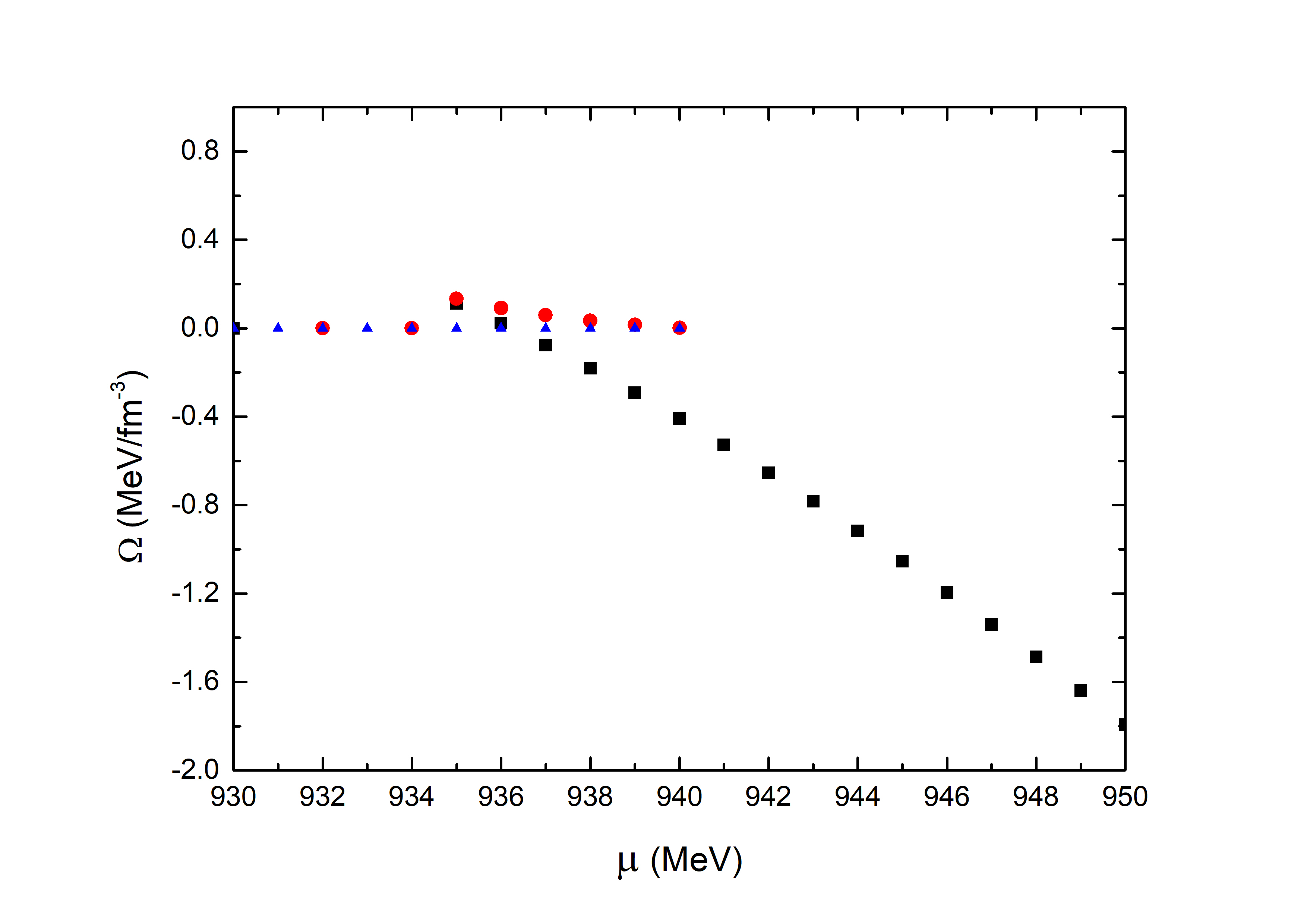}\\
	\caption{The thermodynamical potential as a function of the chemical potential under electric and chemical equilibrium.}\label{fig:potentialE}
\end{figure}

\begin{figure}[H]
	\centering
	\includegraphics[width=1\linewidth]{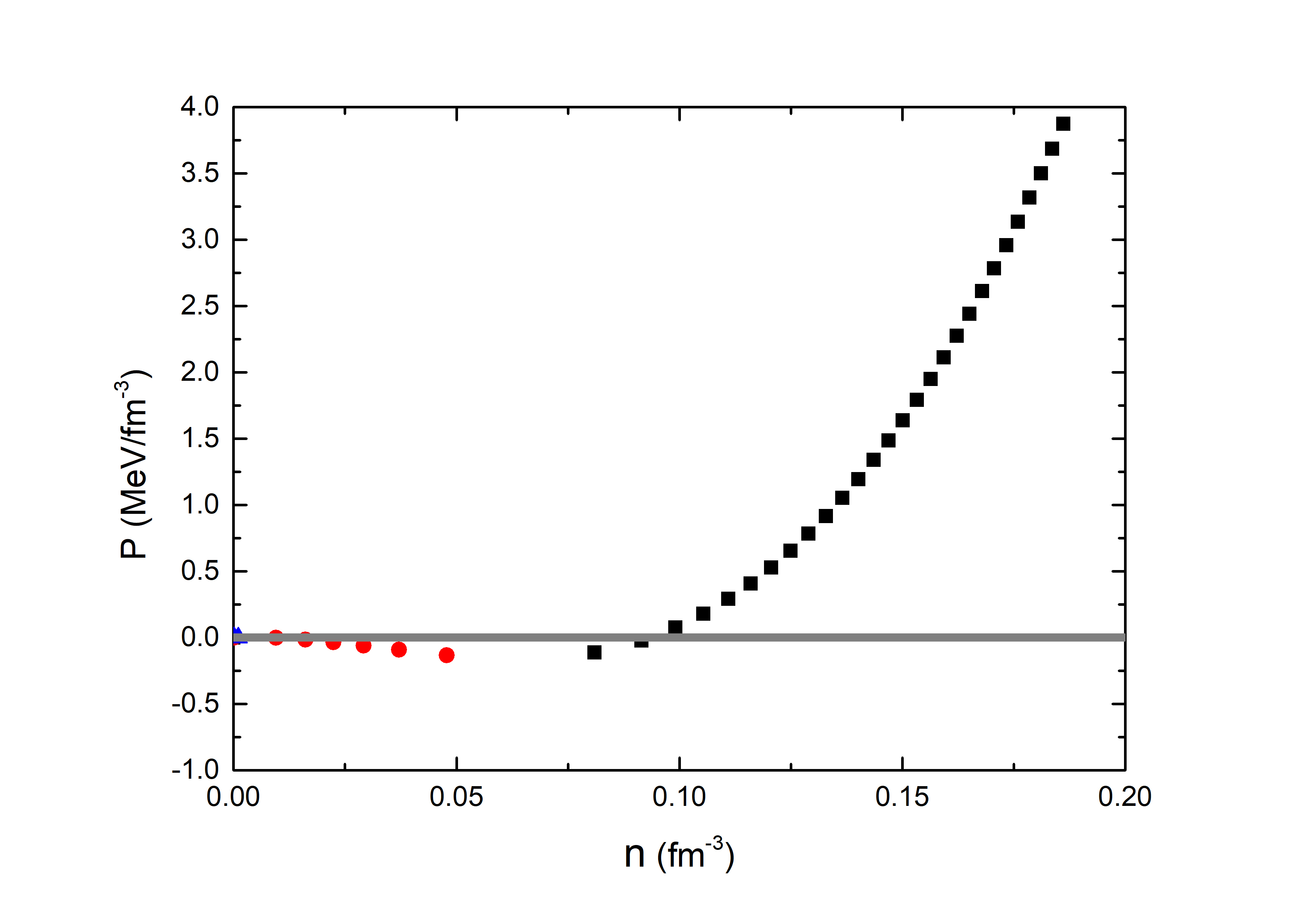}\\
	\caption{The pressure as a function of the baryon number density with different initial values. The grey line corresponds to zero pressure. }
    \label{fig:pressureE}
\end{figure}

In order to describe the hadronic matter in hybrid stars, the electronic and chemical equilibriums
should be included,
\begin{align}
&\mu_p+\mu_e=\mu_n, \\
&\mu_p+\mu_\mu=\mu_n, \\
&n_p-n_e-n_\mu=0.
\end{align}
As shown in Fig. \ref{fig:potentialE} and  Fig. \ref{fig:pressureE}, the thermodynamical properties
under electronic and chemical equilibriums are similar to the case without such conditions, but the
zero pressure point moves from $\mu_0=923~ \mathrm{MeV}$ to $\mu_0=937 ~\mathrm{MeV}$. Correspondingly,
the baryon number density moves from $n_0=0.155 ~\mathrm{fm}^{-3}$ to $\rho_0=0.1 ~\mathrm{fm}^{-3}$.

\section{THE SELF-CONSISTENT NJL MODEL}\label{section:4}

The standard Lagrangian of the two-flavor NJL model \cite{PhysRevD.9.3471,PhysRevD.90.114031,PhysRevD.91.034017,PhysRevD.91.056003,RevModPhys.64.649,MASAYUKI1989668,CLOET20141,ROBERTS2000S1,ROBERTS1994477} with chemical potential is,
\begin{align}
\mathcal{L}=\bar{\psi}(i\gamma^\mu\partial_\mu-m_0+\mu\gamma^0)\psi+G[(\bar{\psi}\psi)^2+(\bar{\psi}i\gamma_5\boldsymbol\tau\psi)^2],\label{eq:1}
\end{align}
where $G$ is the four-fermion coupling constant, $m_0$ is the current quark mass
matrix and$~\boldsymbol\tau~$is the pauli matrix.
The NJL model as an effective field theory can not be solved by the perturbation method. The mean field approximation has to be employed for thermodynamical properties of quark matter \cite{Buballa:2003qv},
\begin{align}
\mathcal{L}_\mathrm{eff}=\bar{\psi}(i\gamma^\mu\partial_\mu-m_0+2G\sigma+\mu\gamma^0)\psi-G\sigma^2,\label{eq:2}
\end{align}
where $~\sigma=\langle\bar{\psi}\psi\rangle~$ is the quark condensate. The effective quark mass is obtained as
\begin{align}
M=m_0-2G\sigma.\label{eq:3}
\end{align}
The quark condensate is derived by setting stationary point of the thermodynamical potential of quark matter with respect to the effective quark mass to be zero:
\begin{align}
\sigma=\langle\bar{\psi}\psi\rangle=\int\dfrac{d^4p}{(2\pi)^4}\mathrm{Tr}[S(p)],\label{eq:4}
\end{align}
where $S(p)$ represents the dressed quark propagator, and the trace operates on the Dirac, flavor and color space. Here it should be noted that the NJL model does not possess the property of confinement. Normally, it is believed that the chiral phase transition and phase transition of deconfinement happen at the same time. Thus, the scalar condensate manifests the chiral phase transition along with the deconfinement.

Since we only concern about the situation of finite chemical potential at zero temperature. The quark condensate is:
\begin{align}
\sigma=-2N_cN_f\int \dfrac{d^3p}{(2\pi)^3}\dfrac{M}{E_p}\big[1-\theta(\mu-E_p)\big],   \label{eq:6}
\end{align}
where $E_p=\sqrt{p^2+M^2}$.
The Fierz-transformation of the four-fermion interaction in the two-flavor NJL model is
\begin{align}
\mathscr{F}=&\dfrac{1}{8\mathrm{N_c}}[2(\bar{\psi}\psi)^2+2(\bar{\psi}i\gamma_5\boldsymbol\tau\psi)^2-2(\bar{\psi}\boldsymbol\tau\psi)^2-2(\bar{\psi}i\gamma_5\psi)^2\nonumber\\
&-4(\bar{\psi}\gamma^{\mu}\psi)^2-4(\bar{\psi}i\gamma^{\mu}\gamma_5\psi)^2+(\bar{\psi}\sigma^{\mu\nu}\psi)^2-(\bar\psi\sigma^{\mu\nu}\boldsymbol\tau\psi)^2],\label{eq:7}
\end{align}
where color terms are neglected. We only keep the scalar
term and the vector term from the Fierz-transformed four-fermion interaction.
As the Fierz transformation is just a mathematical technique,
the original Lagrangian and Fierz-transformed Lagrangian can be combined at any proportions. The weighting factor $\alpha$ reflects the competition between scalar interaction channels and vector interaction channels \cite{Yu:2020dnj,PhysRevD.100.094012,PhysRevD.100.043018,PhysRevD.92.084009,PhysRevD.100.123003,Wang_2019,PhysRevD.102.054028}.
\begin{align}
\mathcal{L}_R= (1-\alpha)\mathcal{L}+\alpha\mathcal{L}_F,\label{eq:8}
\end{align}
where $\mathcal{L}$ is the original Lagrangian and $\mathcal{L}_F$ is Fierz-transformed Lagrangian. However, this situation is different when we employed the mean field approximation, because the Fierz-transformation and mean-field approximation are not commutative:
\begin{align}
&M=m_0-2G(1-\alpha+\dfrac{\alpha}{4N_c})\sigma,\label{eq:9}\\
&\mu'=\mu-\dfrac{G'\alpha}{N_c\pi^2}\langle\psi^\dagger\psi\rangle,   \label{eq:10}
\end{align}
where $~G'=\dfrac{2G(1+\dfrac{1}{4N_c})}{(1-\alpha+\dfrac{\alpha}{4N_c})}~$.
It is clear that the gap equations are affected by $\alpha$ \cite{Wang_2019}, and the vector
interaction between quarks becomes dominant with the increase of $\alpha$. We will show that
the parameter $\alpha$ is crucial to the EOS of quark star, and can be constrained from
astronomical observations.

As indicated by the divergence of quark condensates, the three-momentum cutoff is usually adopted.
But $5\sim10~\mathrm{n}_0$ of the central density of hybrid stars corresponds to quark chemical potential over 700 $\mathrm{MeV}$.
In this region, the $\theta$ function of the equation is malfunctioning,
because the quark chemical potential is beyond the cutoff of the momentum.
We have the mathematical identity
\begin{align}
\dfrac{1}{A^n}=\dfrac{1}{\Gamma(n)}\int_0^{\infty}d\tau \tau^{n-1}e^{-\tau A} \rightarrow \dfrac{1}{\Gamma(n)}\int_{\tau_{UV}}^{\infty}d\tau \tau^{n-1}e^{-\tau A}, \label{eq:11}
\end{align}
where $\tau_{UV}$ is set to regularize the divergence. Thus the proper-time regularization is employed and the quark condensate now can be altered into the form of
\begin{align}
\sigma=-2N_cN_f\Big[\int_{\tau_{UV}}^{\infty}d\tau \dfrac{M}{\tau^2}e^{-\tau E_p^2}-\int \dfrac{d^3p}{(2\pi)^3}\dfrac{M}{E_p}\theta(\mu-E_p)\Big].\label{eq:12}
\end{align}
In our calculations, the parameters will be fixed
as $\mathrm{m_0}=5.0 ~\mathrm{MeV}, ~\mathrm{\tau_{UV}}=1092^{-2}~\mathrm{MeV}$
and ~$\mathrm{G'}=3.086 \times 10^{-6} ~\mathrm{MeV}^{-2}$,
which fit the pion decay constant and pion mass at zero temperature and chemical potential \cite{PhysRevD.100.123003}.

Similar to the hadronic matter, quark matter should be in electronic and chemical equilibrium as well,
\begin{align}
&\mu_u+\mu_e=\mu_d,\\
&\mu_u+\mu_\mu=\mu_d,\\
&\dfrac23 n_u-\dfrac13 n_d - n_e - n_\mu=0.
\end{align}
Due to the asymmetry between u-quark and d-quark in the beta equilibrium system,
the pressure of quark matter is given as \cite{PhysRevD.78.054001,Zong:2008zzb}:
\begin{align}
P(\mu_u, \mu_d)=&P(\mu_u=0,\mu_d=0)+\int_0^{\mu_u} \rho_u(\tilde{\mu}_u,\mu_d=0)d\tilde{\mu}_u\nonumber\\
&+\int_0^{\mu_d} \rho_d(\mu_u,\tilde{\mu}_d)d\tilde{\mu}_d.\label{eq:quarkpressure}
\end{align}
Because no free quarks are observed in nature, following the MIT bag model, the vacuum pressure, $P(\mu_u=0,\mu_d=0)$ in Eq.~(\ref{eq:quarkpressure}), is set to make the pressure of quark matter appears at 2 times of nuclear saturation density $2n_0$, where the deconfinement happens,
\begin{align}
P(\mu=0)=-\int_0^{\mu(2n_0)} \rho d\tilde{\mu},\label{eq:17}
\end{align}
where $\mu(2n_0)=1029~\mathrm{MeV}$, which is derived from the Walecka model.

\section{hybrid stars}\label{section:5}

\begin{figure}[H]
	\centering
	\includegraphics[width=1\linewidth]{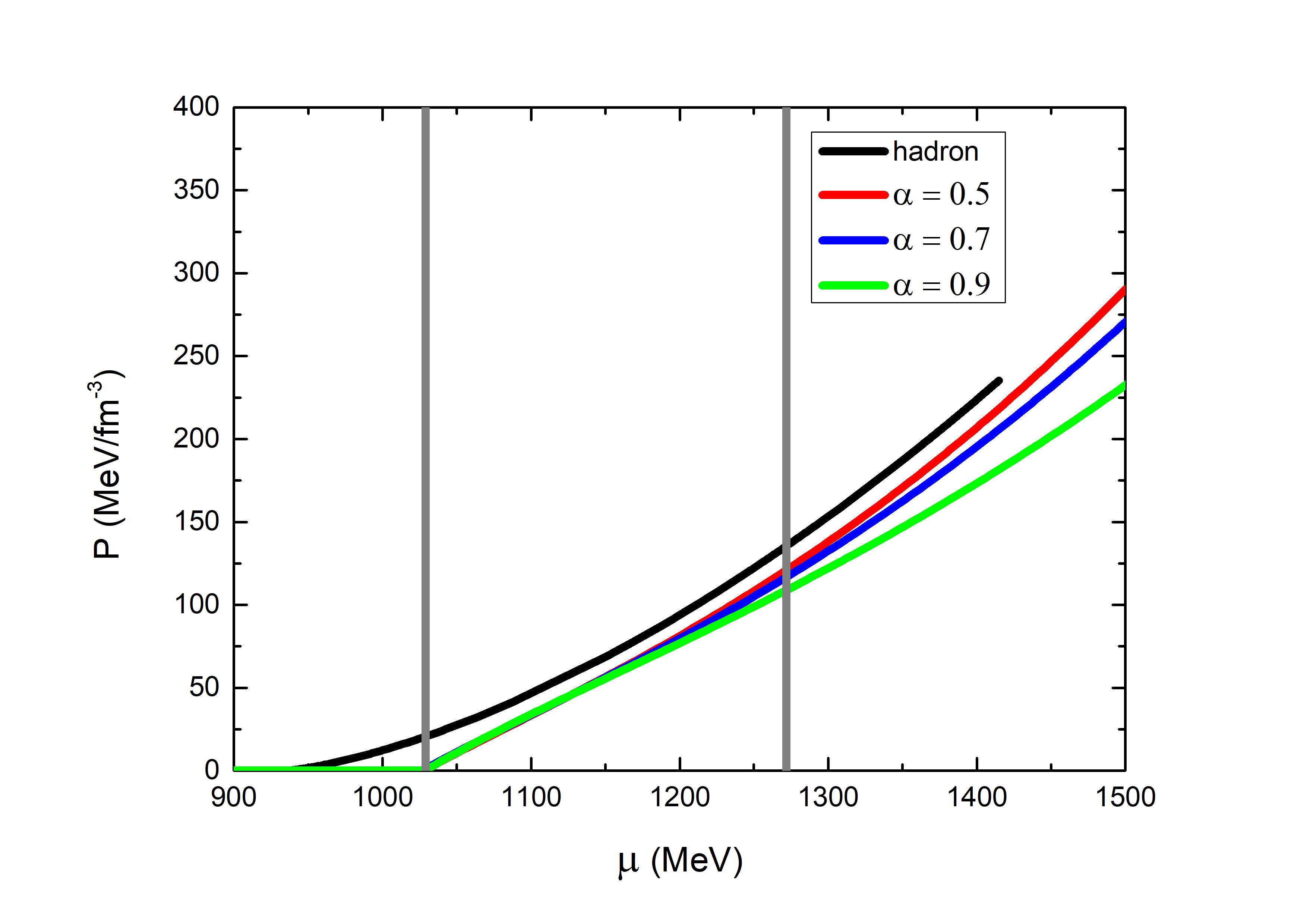}\\
	\caption{The pressure as a function of the chemical potential with different $\alpha$. 
     The black line is the pressure of hadronic matter and the colorful lines are for quark matter. 
     The two vertical grey lines represent the window between 1029 MeV and 1272 MeV, corresponding to $2n_0$ and $4n_0$, where deconfinement happens. }
     \label{fig:threewindow}
\end{figure}

To obtain a smooth EOS of hybrid star, an interpolation approach to connect the hadronic matter at low densities and quark matter at large densities is employed. In Refs. \cite{PhysRevC.93.035807,KOJO2016821,10.1093/ptep/ptt045}, the P-interpolation and $\epsilon$-interpolation approaches in the P-$\rho$ and $\epsilon$-$\rho$ plane are adopted. In this paper, the P-interpolation on the P-$\mu$ plane is employed with an extra parameter $\alpha_I$ introduced as in Refs. \cite{PhysRevD.98.083013,PhysRevD.97.103013,PhysRevD.92.054012},
\begin{align}
&P(\mu)=P_H(\mu)f_-(\mu)+P_Q(\mu)f_+(\mu),\\
&f_\pm=\dfrac12(1\pm \mathrm{tanh}(\dfrac{\mu-\tilde{\mu}}{\added{\alpha_I}\Gamma})).\label{eq:interpolation}
\end{align}
Here, $P_H$ and $P_Q$ represent the pressures of hadronic matter and quark matter respectively. The interpolation functions $f_\pm$ are introduced to make the region of deconfined phase transition smoothly shifting. It is usually assumed that the deconfinement happens at $2\mathrm{n}_0$ and finishes at $4\sim7\mathrm{n}_0$ \cite{Baym_2018}. So, the region of window are set from $2\mathrm{n}_0$ to $4\mathrm{n}_0$, $\tilde{\mu}-\Gamma\leq\mu\leq\tilde{\mu}+\Gamma$ (Fig. \ref{fig:threewindow}), where $\tilde{\mu}$ and $\Gamma$ are  set as the 
chemical potential corresponding to the particle number density at 3 and 1 times of nuclear saturation density respectively, for the reason that no single EOS of hadronic matter or quark matter alone is reliable to describe the deconfinement. Taking into account the thermodynamical relation of Eq.(\ref{eq:thermo relation}), the energy density in such a region is
\begin{align}
&\epsilon(\mu)=\epsilon_H(\mu)f_-(\mu)+\epsilon_Q(\mu)f_+(\mu)+\Delta\epsilon,\\
&\Delta\epsilon=\mu(P_Q-P_H)g(\mu),
\end{align}
where $g(\mu)=\dfrac{2}{\alpha_I\Gamma}(e^Y+e^{-Y})^{-2}$ and $Y=(\mu-\tilde\mu)/(\alpha_I\Gamma)$.

\begin{figure}[H]
	\centering
	\includegraphics[width=1\linewidth]{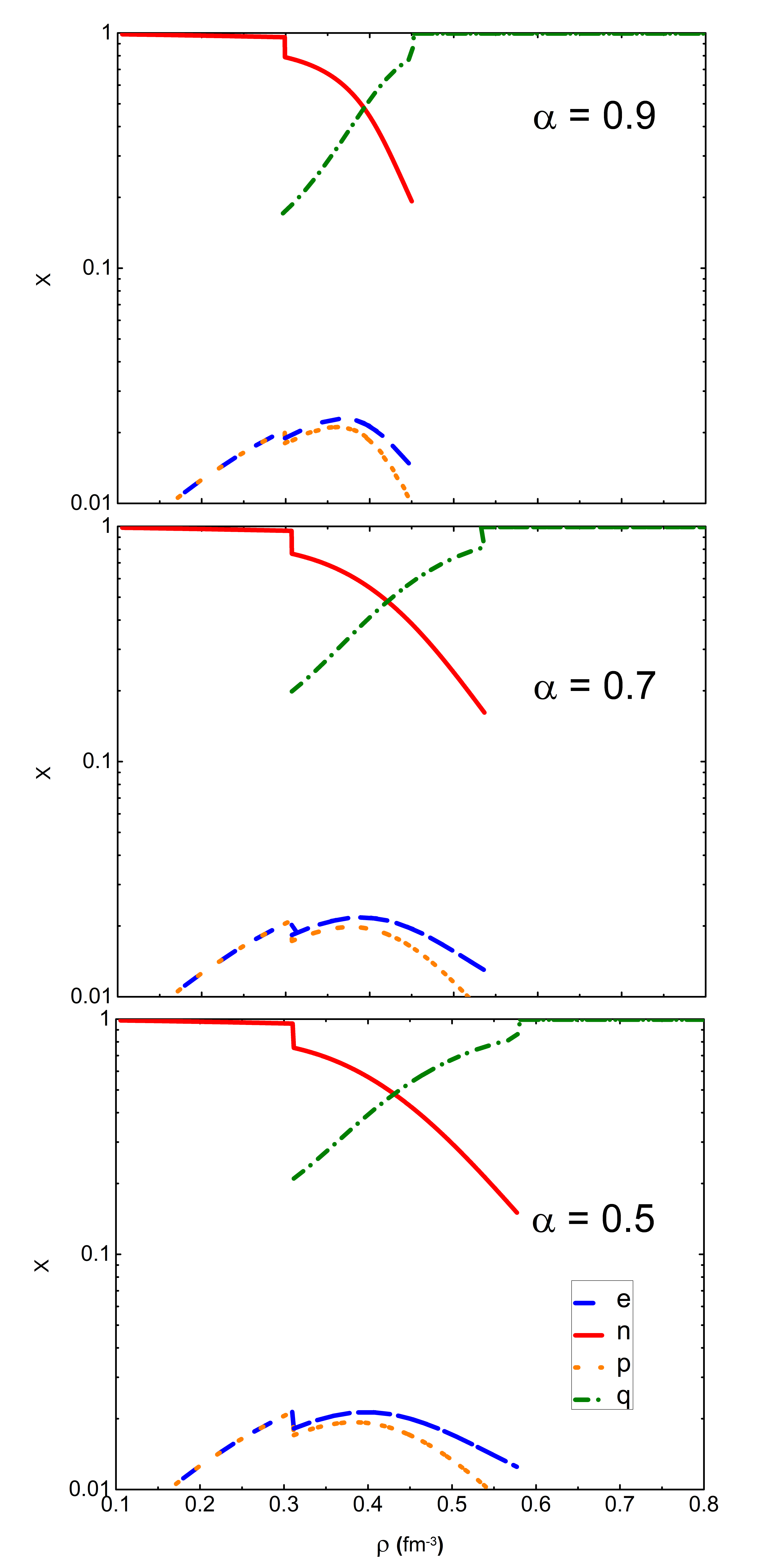}\\
	\caption{The particle fraction of protons (p), neutrons (n), electrons (e) and quarks (q) as a function of the baryon number density. }\label{fig:X}
\end{figure}

With the above interpolation function, the constituent fractions for neutron,  proton, electron and quark are defined as
\begin{align}
X(i) = \dfrac{\rho_i}{\rho_n + \rho_p + \rho_e + \rho_\mu+ \rho_q},~ i=n, p,e,\mu,q \label{eq:X}
\end{align}
where $\rho$ is the baryon or lepton number density and the subscripts of n, p, e, $\mu$ and q stand for neutrons, protons, electrons, muons 
and quarks respectively. The corresponding results for different $\alpha$ 's are shown in Fig. \ref{fig:X}. As the density increases, hadronic matter transforms into quark matter.  Note the muon is also present, but  its fractions are all below 1\%, hence not plotted. (The muon number density is generally suppressed as compared to electron as leptons are treated as free fermion gas here.)
The discontituities in these curves are artifacts of our interpolation scheme. The good thing is that the jumps at these discontinuities are indeed insignificant in magnitude.
$\alpha_I = 0.17$ is assumed in Eq.~(\ref{eq:interpolation}), 
so the hadronic matter fraction declines from 0.8 to 0.2. If $\alpha_I$ decreases further, 
the tidal deformability of the 1.4-solar mass compact star will exceed the upper limit 
from astronomical observations \cite{PhysRevLett.119.161101}. 

\begin{figure}[H]
	\centering
	\includegraphics[width=1\linewidth]{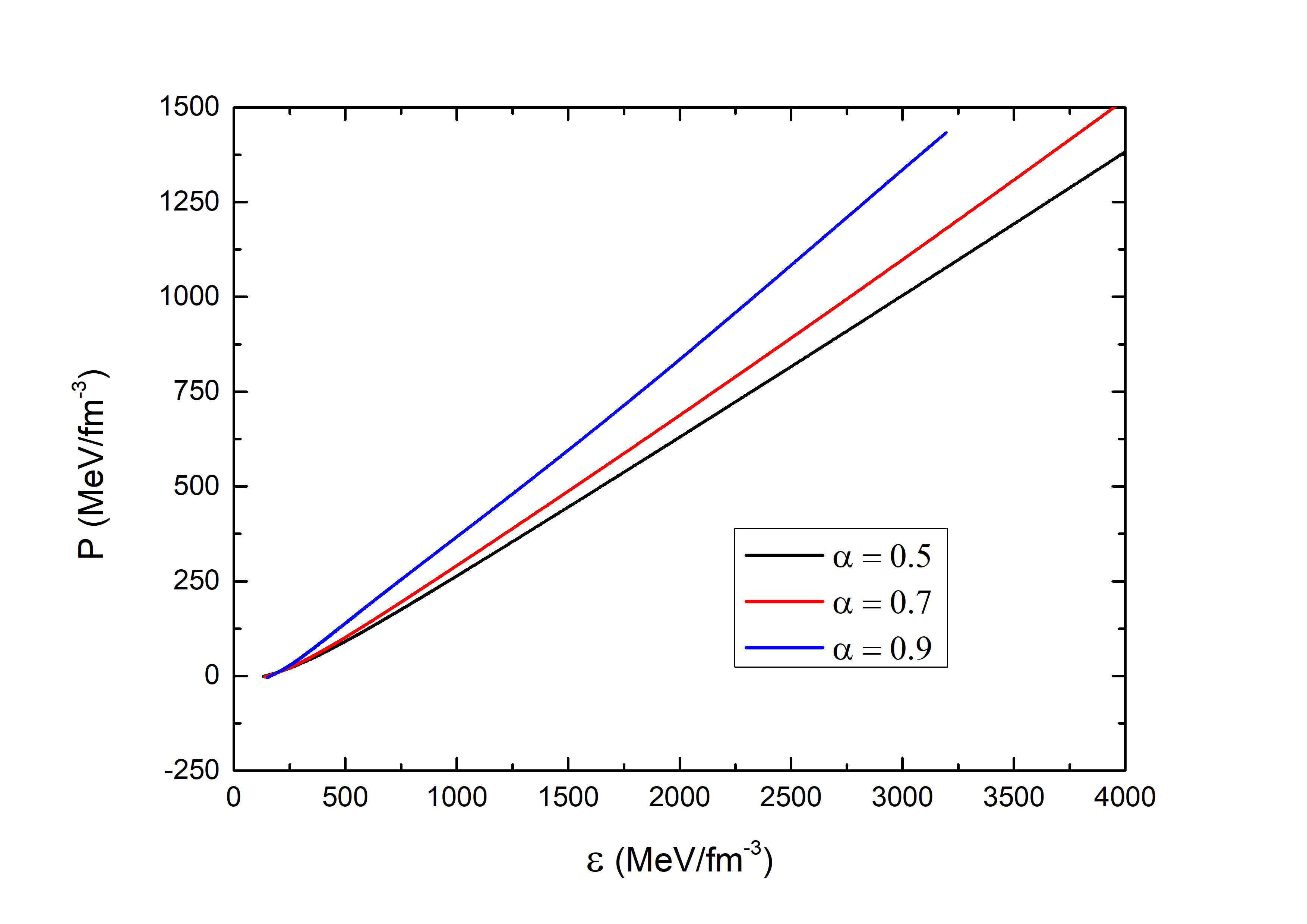}\\
	\caption{The pressure as a function of the energy density for different $\alpha$. }\label{fig:PE}
\end{figure}

By employing such an interpolation, the overall EOS covering all the density ranges can be calculated. 
The result is shown in Fig. \ref{fig:PE}.

\begin{figure}[H]
	\centering
	\includegraphics[width=1\linewidth]{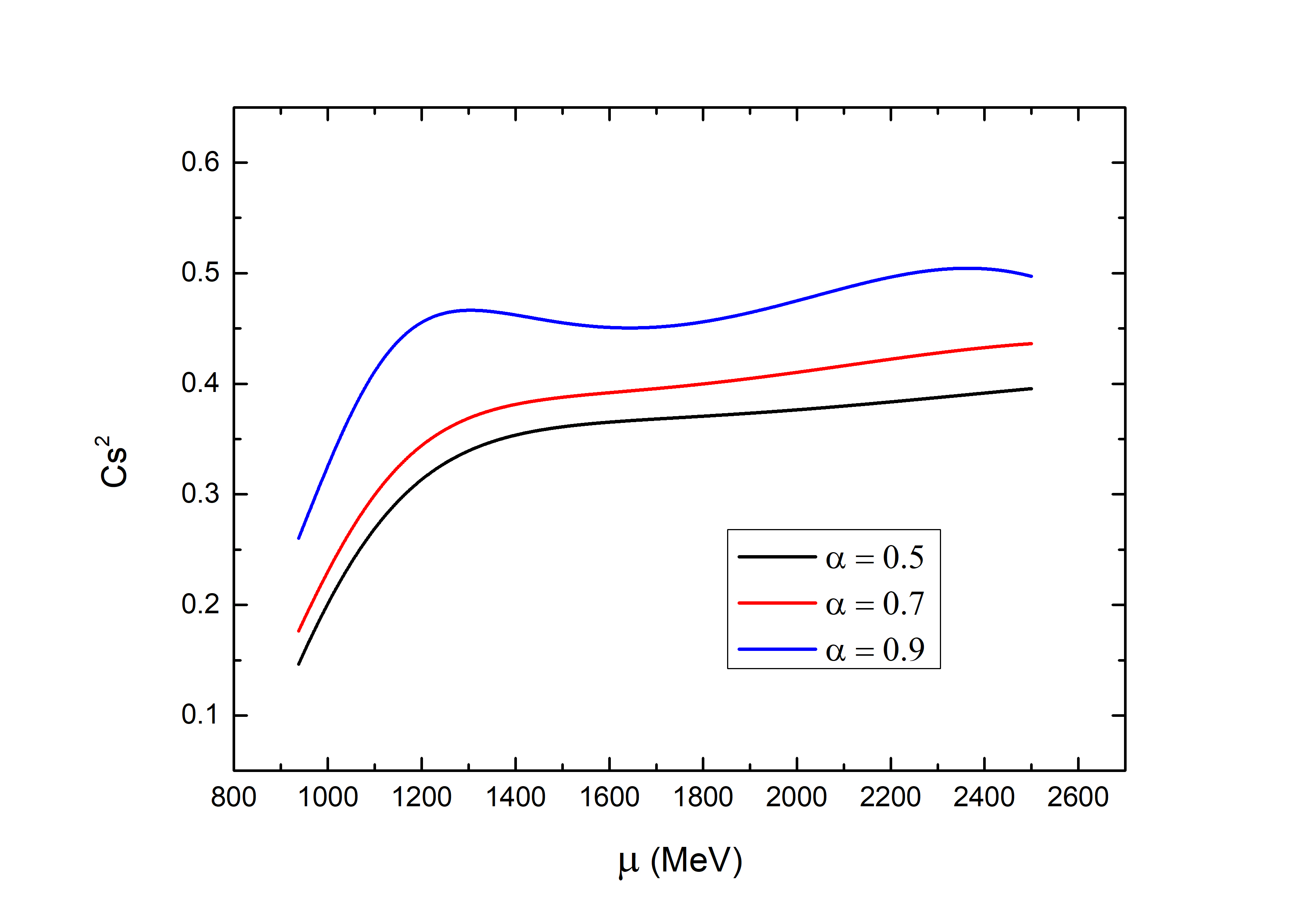}\\
	\caption{The velocity of sound as a function of the chemical potential for different $\alpha$. }\label{fig:Cs}
\end{figure}

The sound velocity can be calculated from the EOS as, 
\begin{align}
C_s^2=\dfrac{dP}{d\epsilon}. 
\end{align}
The sound speed can reflect the stiffness of the EOS. The results are illustrated in Fig.~\ref{fig:Cs}. 
We see that as $\alpha$ grows, the velocity also becomes larger, which means a stiffer EOS. 
Therefore, a stronger vector interaction can support a larger pressure.

Now that the EOS of hybrid stars has been achieved, the Tolman-Oppenheimer-Volkoff equations (TOV) 
can be adopted (as G = c = 1) to solve the structure of the compact star:
\begin{align}
&\dfrac{dP}{dr}=-\dfrac{(\varepsilon+P)(M+4\pi r^3 P)}{r(r-2M)},\label{eq:25}\\
&\dfrac{dM}{dr}=4\pi r^2\varepsilon.\label{eq:26}
\end{align}

\begin{figure}[H]
	\centering
	\includegraphics[width=1\linewidth]{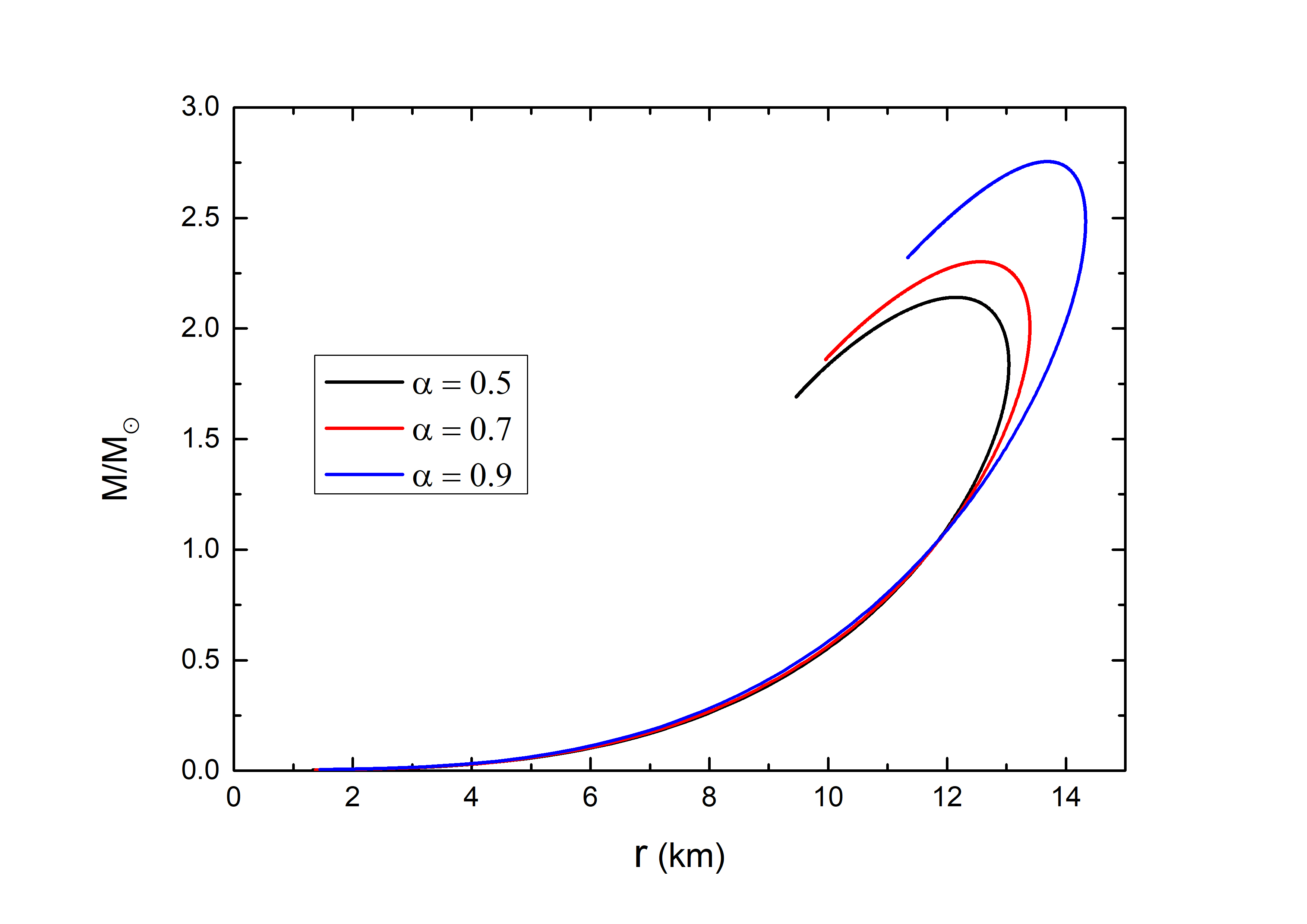}\\
	\caption{The mass-radius relation for different $\alpha$. }\label{fig:MR}
\end{figure}

\begin{figure}[H]
	\centering
	\includegraphics[width=1\linewidth]{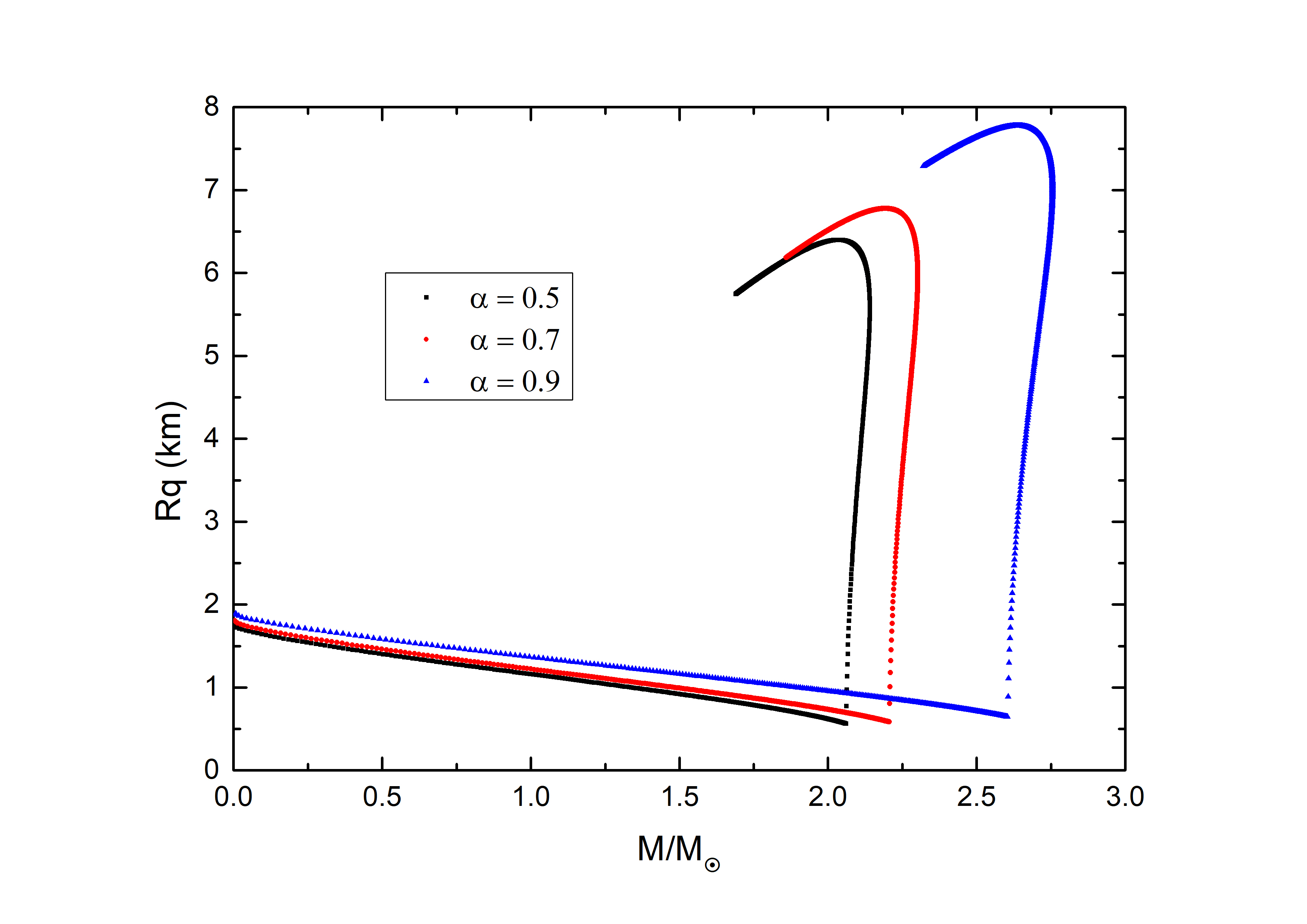}\\
	\caption{The radius of the quark core as a function of mass for hybrid stars. }\label{fig:quarkcore}
\end{figure}

Fig. \ref{fig:MR} shows that, for a larger $\alpha$, the hybrid star can be significantly more 
massive than 2-solar-mass. Unlike the usual mass-radius relation of hybrid stars, for which a small 
mass object usually has a large radius, our hybrid stars with uniform nuclear matter on the surface 
are self-bound, therefore the radius of the hybrid star shrinks as the mass decreases.

Additionally, as demonstrated in Ref. \cite{nature}, a 2-solar-mass neutron star may 
have a 6.5 km quark core, with the square of sound velocity $Cs^2$ lower than 1/3. 
In Fig. \ref{fig:quarkcore}, the radius of quark core ($R_q$) is plot. For the 2-solar-mass 
hybrid star, the radius of quark core is 6.38 km when $\alpha=0.5$, and it is 6.51 km when $\alpha=0.7$. 
The results are compatible with the constraints in Ref. \cite{nature}. However, the square of the 
sound velocity in the NJL model with vector interacting channel is larger than 1/3, which requires 
further study in the future.

The tidal deformability can be calculated with the tidal Love number $k_2$,
and the Love number measures the distortion of the surface of a star by an external gravity.
In the unit $G=c=1$, the relation between $k_2$ and the tidal deformability is 
\begin{align}
k_2=\dfrac{3}{2}\Lambda(\dfrac{M}{R})^5.\label{eq:27}
\end{align}
The $l=2$ tidal Love number $k_2$ is calculated as \cite{PhysRevD.81.123016}
\begin{align}
k_2=&\dfrac{8}{5}C^5(1-2C)^2\big[2+2C(y-1)-y\big]\nonumber\\
&\times \big\{2C[6-3y+3C(5y-8)]\nonumber\\
&+4C^3[13-11y+C(3y-2)+2C^2(1+y)]\nonumber\\
&+3(1-2C)^2[2-y+2C(y-1)]ln(1-2C)\big\}^{-1},\label{eq:28}
\end{align}
where $C=M/R$ is the compactness of the quark star and
\begin{align}
y=\dfrac{R\beta(R)}{H(R)}-\dfrac{4\pi R^3 \varepsilon_0}{M}.    \label{eq:29}
\end{align}
Here $\varepsilon_0$ represents the surface energy density of the quark star.

The dimentionless parameter $y$ is obtained by solving two differential equations
\begin{align}
&~~~~~~~~~~~~~~~\dfrac{dH}{dr}=\beta\label{eq:30},  \\
\dfrac{d\beta}{dr}=&\dfrac{2H}{1-2M/r}\big\{-2\pi\big[5\varepsilon+9P + d\varepsilon/dP(\varepsilon+P)\big]\nonumber\\
&+\dfrac{3}{r^2}+\dfrac{2}{1-2M/r}(\dfrac{M}{r^2}+4\pi rP)^2\big\}\nonumber\\
&+\dfrac{2\beta}{r-2M}[-1+\dfrac{M}{r}+2\pi r^2(\varepsilon-P)].\label{eq:31}
\end{align}
As r $\rightarrow  0$, $H(r)=a_0 r^2$ and $\beta(r)=2a_0 r$. $a_0$ can be any number,
because we only concern about the ratio between H and $\beta$.

\begin{figure}[H]
	\centering
	\includegraphics[width=1\linewidth]{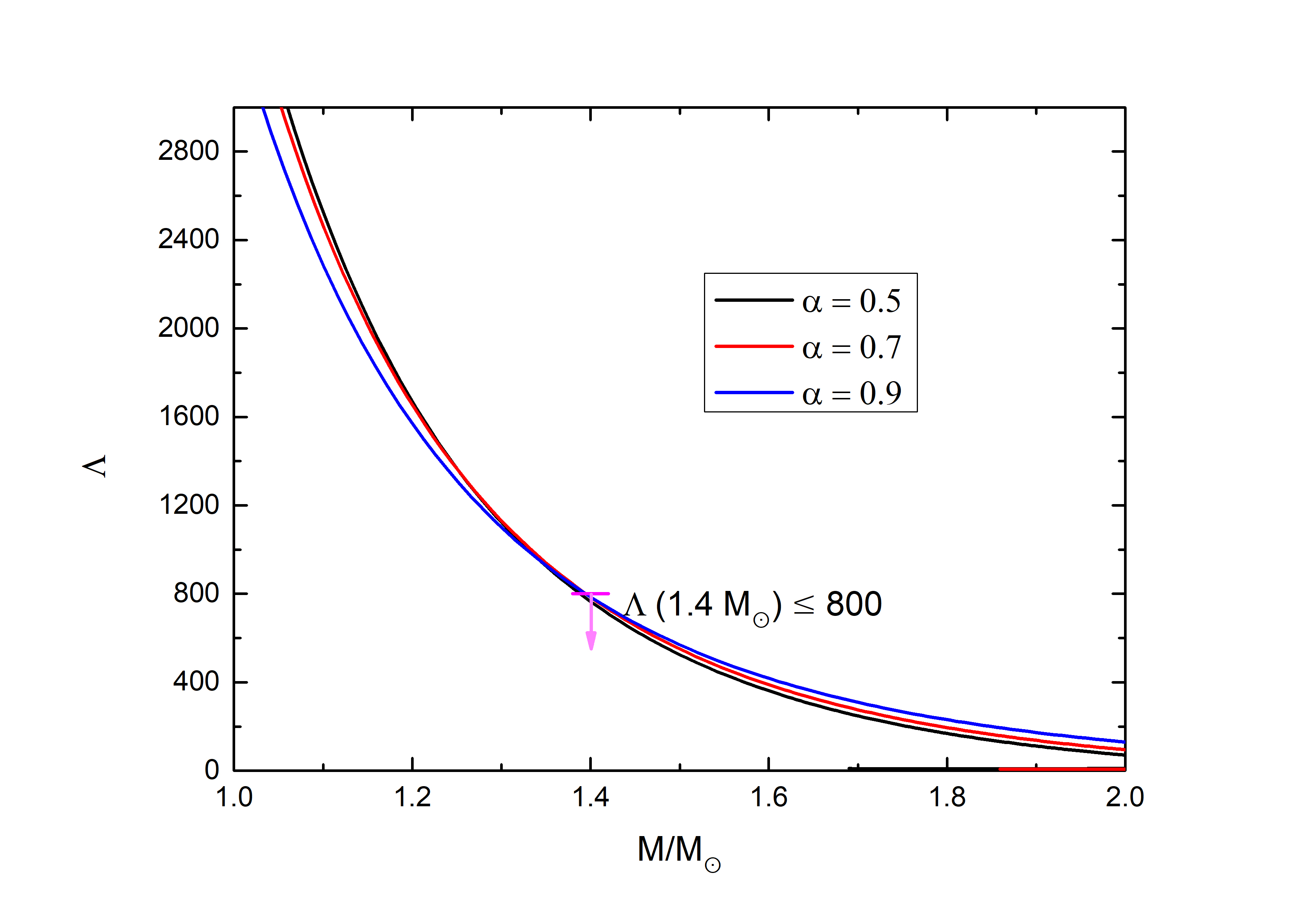}\\
	\caption{Tidal deformability.}\label{fig:Lam}
\end{figure}

In Fig. \ref{fig:Lam}, the tidal deformability $\Lambda$ versus the stellar mass is plot. 
We see that the tidal deformability decreases as the mass of the hybrid star increases. 
The astronomical constraint of $\Lambda<800$ ( \cite{PhysRevLett.119.161101} ) can be 
satisfied as long as $\alpha$ is smaller than 0.9. Hence, we conclude that hybrid stars 
with a uniform surface of two-flavor quark matter can satisfy the current constraints 
from the astronomical observations.

\section{summary and conclusion}\label{section:6}
In this paper, the stability condition of the hadronic matter is discussed. It is pointed out that the most stable state has minimum energy per baryon, $(E/A)_\mathrm{min}=\mu_0$, where nuclear saturation density $\mathrm{n}_0$ is zero pressure. It implies chemical potentials lower than $\mu_0$ is unstable for particles. As for the fact that the iron nucleus is the most stable state, the bulk effect takes responsibility. But in hybrid stars, the bulk effect may not be inappreciable. In addition, unlike a nucleus in a lattice, where pressure appears along with the density, uniform hadronic matter exists at zero pressure. If the inner layer of the hybrid star is the uniform hadronic matter, which is described by the Walecka model, the matter of nuclei is not necessary to be the crust over them. Because, in the description of a nucleus in a lattice, zero pressure must be zero densities, and density at $n\approx\rho_0$ has finite pressures. However, the hadronic matter has no pressure at $n\approx\rho_0$, which makes two phases unable to connect. As a result, hybrid stars or neutron stars with uniform hadronic matter on the surface becomes possible to exist in the universe. For precise explanation, the Walecka model and self-consistent NJL model are employed to construct the hybrid stars, and P-interpolation is used to smoothly connect the EOSs of uniform hadronic matter and quark matter. The mass-radius relations are obtained, which is self-bound rather than gravitationally bound. The astronomical observations provides that two-solar-mass neutron stars exist and 1.4-solar-mass neutron stars has upper limit of tidal deformability $\Lambda<800$, which are all satisfied by our model.

\section*{Acknowledgements}
This work is supported in part by National SKA Program of China No. 2020SKA0120300,
by the National Natural Science Foundation of China (under Grants No. 11475085,
No. 11535005, No. 11905104, No. 11690030, No. 11873030, No. 12041306, and No. U1938201),
by Nation Major State Basic Research and Development of China (2016YFE0129300),
and by the Strategic Priority Research Program of the Chinese Academy of
Sciences (``multi-waveband Gravitational-Wave Universe'', Grant No. XDB23040000).

\bibliography{ref}

\begin{thebibliography}{57}%
\makeatletter
\providecommand \@ifxundefined [1]{%
 \@ifx{#1\undefined}
}%
\providecommand \@ifnum [1]{%
 \ifnum #1\expandafter \@firstoftwo
 \else \expandafter \@secondoftwo
 \fi
}%
\providecommand \@ifx [1]{%
 \ifx #1\expandafter \@firstoftwo
 \else \expandafter \@secondoftwo
 \fi
}%
\providecommand \natexlab [1]{#1}%
\providecommand \enquote  [1]{``#1''}%
\providecommand \bibnamefont  [1]{#1}%
\providecommand \bibfnamefont [1]{#1}%
\providecommand \citenamefont [1]{#1}%
\providecommand \href@noop [0]{\@secondoftwo}%
\providecommand \href [0]{\begingroup \@sanitize@url \@href}%
\providecommand \@href[1]{\@@startlink{#1}\@@href}%
\providecommand \@@href[1]{\endgroup#1\@@endlink}%
\providecommand \@sanitize@url [0]{\catcode `\\12\catcode `\$12\catcode
  `\&12\catcode `\#12\catcode `\^12\catcode `\_12\catcode `\%12\relax}%
\providecommand \@@startlink[1]{}%
\providecommand \@@endlink[0]{}%
\providecommand \url  [0]{\begingroup\@sanitize@url \@url }%
\providecommand \@url [1]{\endgroup\@href {#1}{\urlprefix }}%
\providecommand \urlprefix  [0]{URL }%
\providecommand \Eprint [0]{\href }%
\providecommand \doibase [0]{http://dx.doi.org/}%
\providecommand \selectlanguage [0]{\@gobble}%
\providecommand \bibinfo  [0]{\@secondoftwo}%
\providecommand \bibfield  [0]{\@secondoftwo}%
\providecommand \translation [1]{[#1]}%
\providecommand \BibitemOpen [0]{}%
\providecommand \bibitemStop [0]{}%
\providecommand \bibitemNoStop [0]{.\EOS\space}%
\providecommand \EOS [0]{\spacefactor3000\relax}%
\providecommand \BibitemShut  [1]{\csname bibitem#1\endcsname}%
\let\auto@bib@innerbib\@empty
\bibitem [{\citenamefont {Glendenning}(2012)}]{glendenning2012compact}%
  \BibitemOpen
  \bibfield  {author} {\bibinfo {author} {\bibfnamefont {N.}~\bibnamefont
  {Glendenning}},\ }\href {https://books.google.com/books?id=cCDlBwAAQBAJ}
  {\emph {\bibinfo {title} {Compact Stars: Nuclear Physics, Particle Physics
  and General Relativity}}},\ Astronomy and Astrophysics Library\ (\bibinfo
  {publisher} {Springer New York},\ \bibinfo {year} {2012})\BibitemShut
  {NoStop}%
\bibitem [{\citenamefont {M\"uller}\ and\ \citenamefont
  {Serot}(1995)}]{PhysRevC.52.2072}%
  \BibitemOpen
  \bibfield  {author} {\bibinfo {author} {\bibfnamefont {H.}~\bibnamefont
  {M\"uller}}\ and\ \bibinfo {author} {\bibfnamefont {B.~D.}\ \bibnamefont
  {Serot}},\ }\href {\doibase 10.1103/PhysRevC.52.2072} {\bibfield  {journal}
  {\bibinfo  {journal} {Phys. Rev. C}\ }\textbf {\bibinfo {volume} {52}},\
  \bibinfo {pages} {2072} (\bibinfo {year} {1995})}\BibitemShut {NoStop}%
\bibitem [{\citenamefont {Baym}\ \emph {et~al.}(2018)\citenamefont {Baym},
  \citenamefont {Hatsuda}, \citenamefont {Kojo}, \citenamefont {Powell},
  \citenamefont {Song},\ and\ \citenamefont {Takatsuka}}]{Baym_2018}%
  \BibitemOpen
  \bibfield  {author} {\bibinfo {author} {\bibfnamefont {G.}~\bibnamefont
  {Baym}}, \bibinfo {author} {\bibfnamefont {T.}~\bibnamefont {Hatsuda}},
  \bibinfo {author} {\bibfnamefont {T.}~\bibnamefont {Kojo}}, \bibinfo {author}
  {\bibfnamefont {P.~D.}\ \bibnamefont {Powell}}, \bibinfo {author}
  {\bibfnamefont {Y.}~\bibnamefont {Song}}, \ and\ \bibinfo {author}
  {\bibfnamefont {T.}~\bibnamefont {Takatsuka}},\ }\href {\doibase
  10.1088/1361-6633/aaae14} {\bibfield  {journal} {\bibinfo  {journal} {Reports
  on Progress in Physics}\ }\textbf {\bibinfo {volume} {81}},\ \bibinfo {pages}
  {056902} (\bibinfo {year} {2018})}\BibitemShut {NoStop}%
\bibitem [{\citenamefont {Serot}\ and\ \citenamefont
  {Walecka}(1986{\natexlab{a}})}]{Serot:1984ey}%
  \BibitemOpen
  \bibfield  {author} {\bibinfo {author} {\bibfnamefont {B.~D.}\ \bibnamefont
  {Serot}}\ and\ \bibinfo {author} {\bibfnamefont {J.~D.}\ \bibnamefont
  {Walecka}},\ }\href@noop {} {\bibfield  {journal} {\bibinfo  {journal} {Adv.
  Nucl. Phys.}\ }\textbf {\bibinfo {volume} {16}},\ \bibinfo {pages} {1}
  (\bibinfo {year} {1986}{\natexlab{a}})}\BibitemShut {NoStop}%
\bibitem [{\citenamefont {Avancini}\ \emph {et~al.}(2008)\citenamefont
  {Avancini}, \citenamefont {Menezes}, \citenamefont {Alloy}, \citenamefont
  {Marinelli}, \citenamefont {Moraes},\ and\ \citenamefont
  {Provid\^encia}}]{PhysRevC.78.015802}%
  \BibitemOpen
  \bibfield  {author} {\bibinfo {author} {\bibfnamefont {S.~S.}\ \bibnamefont
  {Avancini}}, \bibinfo {author} {\bibfnamefont {D.~P.}\ \bibnamefont
  {Menezes}}, \bibinfo {author} {\bibfnamefont {M.~D.}\ \bibnamefont {Alloy}},
  \bibinfo {author} {\bibfnamefont {J.~R.}\ \bibnamefont {Marinelli}}, \bibinfo
  {author} {\bibfnamefont {M.~M.~W.}\ \bibnamefont {Moraes}}, \ and\ \bibinfo
  {author} {\bibfnamefont {C.}~\bibnamefont {Provid\^encia}},\ }\href {\doibase
  10.1103/PhysRevC.78.015802} {\bibfield  {journal} {\bibinfo  {journal} {Phys.
  Rev. C}\ }\textbf {\bibinfo {volume} {78}},\ \bibinfo {pages} {015802}
  (\bibinfo {year} {2008})}\BibitemShut {NoStop}%
\bibitem [{\citenamefont {Okamoto}\ \emph {et~al.}(2013)\citenamefont
  {Okamoto}, \citenamefont {Maruyama}, \citenamefont {Yabana},\ and\
  \citenamefont {Tatsumi}}]{PhysRevC.88.025801}%
  \BibitemOpen
  \bibfield  {author} {\bibinfo {author} {\bibfnamefont {M.}~\bibnamefont
  {Okamoto}}, \bibinfo {author} {\bibfnamefont {T.}~\bibnamefont {Maruyama}},
  \bibinfo {author} {\bibfnamefont {K.}~\bibnamefont {Yabana}}, \ and\ \bibinfo
  {author} {\bibfnamefont {T.}~\bibnamefont {Tatsumi}},\ }\href {\doibase
  10.1103/PhysRevC.88.025801} {\bibfield  {journal} {\bibinfo  {journal} {Phys.
  Rev. C}\ }\textbf {\bibinfo {volume} {88}},\ \bibinfo {pages} {025801}
  (\bibinfo {year} {2013})}\BibitemShut {NoStop}%
\bibitem [{\citenamefont {Chamel}\ and\ \citenamefont
  {Haensel}(2008)}]{Chamel:2008ca}%
  \BibitemOpen
  \bibfield  {author} {\bibinfo {author} {\bibfnamefont {N.}~\bibnamefont
  {Chamel}}\ and\ \bibinfo {author} {\bibfnamefont {P.}~\bibnamefont
  {Haensel}},\ }\href {\doibase 10.12942/lrr-2008-10} {\bibfield  {journal}
  {\bibinfo  {journal} {Living Rev. Rel.}\ }\textbf {\bibinfo {volume} {11}},\
  \bibinfo {pages} {10} (\bibinfo {year} {2008})},\ \Eprint
  {http://arxiv.org/abs/0812.3955} {arXiv:0812.3955 [astro-ph]} \BibitemShut
  {NoStop}%
\bibitem [{\citenamefont {Fortin}\ \emph {et~al.}(2016)\citenamefont {Fortin},
  \citenamefont {Provid\^encia}, \citenamefont {Raduta}, \citenamefont
  {Gulminelli}, \citenamefont {Zdunik}, \citenamefont {Haensel},\ and\
  \citenamefont {Bejger}}]{PhysRevC.94.035804}%
  \BibitemOpen
  \bibfield  {author} {\bibinfo {author} {\bibfnamefont {M.}~\bibnamefont
  {Fortin}}, \bibinfo {author} {\bibfnamefont {C.}~\bibnamefont
  {Provid\^encia}}, \bibinfo {author} {\bibfnamefont {A.~R.}\ \bibnamefont
  {Raduta}}, \bibinfo {author} {\bibfnamefont {F.}~\bibnamefont {Gulminelli}},
  \bibinfo {author} {\bibfnamefont {J.~L.}\ \bibnamefont {Zdunik}}, \bibinfo
  {author} {\bibfnamefont {P.}~\bibnamefont {Haensel}}, \ and\ \bibinfo
  {author} {\bibfnamefont {M.}~\bibnamefont {Bejger}},\ }\href {\doibase
  10.1103/PhysRevC.94.035804} {\bibfield  {journal} {\bibinfo  {journal} {Phys.
  Rev. C}\ }\textbf {\bibinfo {volume} {94}},\ \bibinfo {pages} {035804}
  (\bibinfo {year} {2016})}\BibitemShut {NoStop}%
\bibitem [{\citenamefont {Zhu}\ and\ \citenamefont
  {Li}(2018)}]{PhysRevC.97.035805}%
  \BibitemOpen
  \bibfield  {author} {\bibinfo {author} {\bibfnamefont {Z.-Y.}\ \bibnamefont
  {Zhu}}\ and\ \bibinfo {author} {\bibfnamefont {A.}~\bibnamefont {Li}},\
  }\href {\doibase 10.1103/PhysRevC.97.035805} {\bibfield  {journal} {\bibinfo
  {journal} {Phys. Rev. C}\ }\textbf {\bibinfo {volume} {97}},\ \bibinfo
  {pages} {035805} (\bibinfo {year} {2018})}\BibitemShut {NoStop}%
\bibitem [{\citenamefont {Abbott}\ and\ \citenamefont
  {Abbott}(2017)}]{PhysRevLett.119.161101}%
  \BibitemOpen
  \bibfield  {author} {\bibinfo {author} {\bibfnamefont {B.~P.}\ \bibnamefont
  {Abbott}}\ and\ \bibinfo {author} {\bibfnamefont {R.}~\bibnamefont {Abbott}}
  (\bibinfo {collaboration} {LIGO Scientific Collaboration and Virgo
  Collaboration}),\ }\href {\doibase 10.1103/PhysRevLett.119.161101} {\bibfield
   {journal} {\bibinfo  {journal} {Phys. Rev. Lett.}\ }\textbf {\bibinfo
  {volume} {119}},\ \bibinfo {pages} {161101} (\bibinfo {year}
  {2017})}\BibitemShut {NoStop}%
\bibitem [{\citenamefont {{Annala}}\ \emph {et~al.}(2020)\citenamefont
  {{Annala}}, \citenamefont {{Gorda}}, \citenamefont {{Kurkela}}, \citenamefont
  {{N{\"a}ttil{\"a}}},\ and\ \citenamefont {{Vuorinen}}}]{nature}%
  \BibitemOpen
  \bibfield  {author} {\bibinfo {author} {\bibfnamefont {E.}~\bibnamefont
  {{Annala}}}, \bibinfo {author} {\bibfnamefont {T.}~\bibnamefont {{Gorda}}},
  \bibinfo {author} {\bibfnamefont {A.}~\bibnamefont {{Kurkela}}}, \bibinfo
  {author} {\bibfnamefont {J.}~\bibnamefont {{N{\"a}ttil{\"a}}}}, \ and\
  \bibinfo {author} {\bibfnamefont {A.}~\bibnamefont {{Vuorinen}}},\ }\href
  {\doibase 10.1038/s41567-020-0914-9} {\bibfield  {journal} {\bibinfo
  {journal} {Nature Physics}\ }\textbf {\bibinfo {volume} {16}},\ \bibinfo
  {pages} {907} (\bibinfo {year} {2020})},\ \Eprint
  {http://arxiv.org/abs/1903.09121} {arXiv:1903.09121 [astro-ph.HE]}
  \BibitemShut {NoStop}%
\bibitem [{\citenamefont {Antoniadis}\ \emph {et~al.}(2013)\citenamefont
  {Antoniadis}, \citenamefont {Freire}, \citenamefont {Wex}, \citenamefont
  {Tauris}, \citenamefont {Lynch}, \citenamefont {van Kerkwijk}, \citenamefont
  {Kramer}, \citenamefont {Bassa}, \citenamefont {Dhillon}, \citenamefont
  {Driebe}, \citenamefont {Hessels}, \citenamefont {Kaspi}, \citenamefont
  {Kondratiev}, \citenamefont {Langer}, \citenamefont {Marsh}, \citenamefont
  {McLaughlin}, \citenamefont {Pennucci}, \citenamefont {Ransom}, \citenamefont
  {Stairs}, \citenamefont {van Leeuwen}, \citenamefont {Verbiest},\ and\
  \citenamefont {Whelan}}]{Antoniadis1233232}%
  \BibitemOpen
  \bibfield  {author} {\bibinfo {author} {\bibfnamefont {J.}~\bibnamefont
  {Antoniadis}}, \bibinfo {author} {\bibfnamefont {P.~C.~C.}\ \bibnamefont
  {Freire}}, \bibinfo {author} {\bibfnamefont {N.}~\bibnamefont {Wex}},
  \bibinfo {author} {\bibfnamefont {T.~M.}\ \bibnamefont {Tauris}}, \bibinfo
  {author} {\bibfnamefont {R.~S.}\ \bibnamefont {Lynch}}, \bibinfo {author}
  {\bibfnamefont {M.~H.}\ \bibnamefont {van Kerkwijk}}, \bibinfo {author}
  {\bibfnamefont {M.}~\bibnamefont {Kramer}}, \bibinfo {author} {\bibfnamefont
  {C.}~\bibnamefont {Bassa}}, \bibinfo {author} {\bibfnamefont {V.~S.}\
  \bibnamefont {Dhillon}}, \bibinfo {author} {\bibfnamefont {T.}~\bibnamefont
  {Driebe}}, \bibinfo {author} {\bibfnamefont {J.~W.~T.}\ \bibnamefont
  {Hessels}}, \bibinfo {author} {\bibfnamefont {V.~M.}\ \bibnamefont {Kaspi}},
  \bibinfo {author} {\bibfnamefont {V.~I.}\ \bibnamefont {Kondratiev}},
  \bibinfo {author} {\bibfnamefont {N.}~\bibnamefont {Langer}}, \bibinfo
  {author} {\bibfnamefont {T.~R.}\ \bibnamefont {Marsh}}, \bibinfo {author}
  {\bibfnamefont {M.~A.}\ \bibnamefont {McLaughlin}}, \bibinfo {author}
  {\bibfnamefont {T.~T.}\ \bibnamefont {Pennucci}}, \bibinfo {author}
  {\bibfnamefont {S.~M.}\ \bibnamefont {Ransom}}, \bibinfo {author}
  {\bibfnamefont {I.~H.}\ \bibnamefont {Stairs}}, \bibinfo {author}
  {\bibfnamefont {J.}~\bibnamefont {van Leeuwen}}, \bibinfo {author}
  {\bibfnamefont {J.~P.~W.}\ \bibnamefont {Verbiest}}, \ and\ \bibinfo {author}
  {\bibfnamefont {D.~G.}\ \bibnamefont {Whelan}},\ }\href {\doibase
  10.1126/science.1233232} {\bibfield  {journal} {\bibinfo  {journal}
  {Science}\ }\textbf {\bibinfo {volume} {340}} (\bibinfo {year} {2013}),\
  10.1126/science.1233232},\ \Eprint
  {http://arxiv.org/abs/https://science.sciencemag.org/content/340/6131/1233232.full.pdf}
  {https://science.sciencemag.org/content/340/6131/1233232.full.pdf}
  \BibitemShut {NoStop}%
\bibitem [{\citenamefont {Bogdanov}\ \emph {et~al.}(2019)\citenamefont
  {Bogdanov}, \citenamefont {Guillot}, \citenamefont {Ray}, \citenamefont
  {Wolff}, \citenamefont {Chakrabarty}, \citenamefont {Ho}, \citenamefont
  {Kerr}, \citenamefont {Lamb}, \citenamefont {Lommen}, \citenamefont {Ludlam},
  \citenamefont {Milburn}, \citenamefont {Montano}, \citenamefont {Miller},
  \citenamefont {Bauböck}, \citenamefont {Özel}, \citenamefont {Psaltis},
  \citenamefont {Remillard}, \citenamefont {Riley}, \citenamefont {Steiner},
  \citenamefont {Strohmayer}, \citenamefont {Watts}, \citenamefont {Wood},
  \citenamefont {Zeldes}, \citenamefont {Enoto}, \citenamefont {Okajima},
  \citenamefont {Kellogg}, \citenamefont {Baker}, \citenamefont {Markwardt},
  \citenamefont {Arzoumanian},\ and\ \citenamefont {Gendreau}}]{Bogdanov_2019}%
  \BibitemOpen
  \bibfield  {author} {\bibinfo {author} {\bibfnamefont {S.}~\bibnamefont
  {Bogdanov}}, \bibinfo {author} {\bibfnamefont {S.}~\bibnamefont {Guillot}},
  \bibinfo {author} {\bibfnamefont {P.~S.}\ \bibnamefont {Ray}}, \bibinfo
  {author} {\bibfnamefont {M.~T.}\ \bibnamefont {Wolff}}, \bibinfo {author}
  {\bibfnamefont {D.}~\bibnamefont {Chakrabarty}}, \bibinfo {author}
  {\bibfnamefont {W.~C.~G.}\ \bibnamefont {Ho}}, \bibinfo {author}
  {\bibfnamefont {M.}~\bibnamefont {Kerr}}, \bibinfo {author} {\bibfnamefont
  {F.~K.}\ \bibnamefont {Lamb}}, \bibinfo {author} {\bibfnamefont
  {A.}~\bibnamefont {Lommen}}, \bibinfo {author} {\bibfnamefont {R.~M.}\
  \bibnamefont {Ludlam}}, \bibinfo {author} {\bibfnamefont {R.}~\bibnamefont
  {Milburn}}, \bibinfo {author} {\bibfnamefont {S.}~\bibnamefont {Montano}},
  \bibinfo {author} {\bibfnamefont {M.~C.}\ \bibnamefont {Miller}}, \bibinfo
  {author} {\bibfnamefont {M.}~\bibnamefont {Bauböck}}, \bibinfo {author}
  {\bibfnamefont {F.}~\bibnamefont {Özel}}, \bibinfo {author} {\bibfnamefont
  {D.}~\bibnamefont {Psaltis}}, \bibinfo {author} {\bibfnamefont {R.~A.}\
  \bibnamefont {Remillard}}, \bibinfo {author} {\bibfnamefont {T.~E.}\
  \bibnamefont {Riley}}, \bibinfo {author} {\bibfnamefont {J.~F.}\ \bibnamefont
  {Steiner}}, \bibinfo {author} {\bibfnamefont {T.~E.}\ \bibnamefont
  {Strohmayer}}, \bibinfo {author} {\bibfnamefont {A.~L.}\ \bibnamefont
  {Watts}}, \bibinfo {author} {\bibfnamefont {K.~S.}\ \bibnamefont {Wood}},
  \bibinfo {author} {\bibfnamefont {J.}~\bibnamefont {Zeldes}}, \bibinfo
  {author} {\bibfnamefont {T.}~\bibnamefont {Enoto}}, \bibinfo {author}
  {\bibfnamefont {T.}~\bibnamefont {Okajima}}, \bibinfo {author} {\bibfnamefont
  {J.~W.}\ \bibnamefont {Kellogg}}, \bibinfo {author} {\bibfnamefont
  {C.}~\bibnamefont {Baker}}, \bibinfo {author} {\bibfnamefont {C.~B.}\
  \bibnamefont {Markwardt}}, \bibinfo {author} {\bibfnamefont {Z.}~\bibnamefont
  {Arzoumanian}}, \ and\ \bibinfo {author} {\bibfnamefont {K.~C.}\ \bibnamefont
  {Gendreau}},\ }\href {\doibase 10.3847/2041-8213/ab53eb} {\bibfield
  {journal} {\bibinfo  {journal} {The Astrophysical Journal}\ }\textbf
  {\bibinfo {volume} {887}},\ \bibinfo {pages} {L25} (\bibinfo {year}
  {2019})}\BibitemShut {NoStop}%
\bibitem [{\citenamefont {Riley}\ \emph {et~al.}(2019)\citenamefont {Riley},
  \citenamefont {Watts}, \citenamefont {Bogdanov}, \citenamefont {Ray},
  \citenamefont {Ludlam}, \citenamefont {Guillot}, \citenamefont {Arzoumanian},
  \citenamefont {Baker}, \citenamefont {Bilous}, \citenamefont {Chakrabarty},
  \citenamefont {Gendreau}, \citenamefont {Harding}, \citenamefont {Ho},
  \citenamefont {Lattimer}, \citenamefont {Morsink},\ and\ \citenamefont
  {Strohmayer}}]{Riley_2019}%
  \BibitemOpen
  \bibfield  {author} {\bibinfo {author} {\bibfnamefont {T.~E.}\ \bibnamefont
  {Riley}}, \bibinfo {author} {\bibfnamefont {A.~L.}\ \bibnamefont {Watts}},
  \bibinfo {author} {\bibfnamefont {S.}~\bibnamefont {Bogdanov}}, \bibinfo
  {author} {\bibfnamefont {P.~S.}\ \bibnamefont {Ray}}, \bibinfo {author}
  {\bibfnamefont {R.~M.}\ \bibnamefont {Ludlam}}, \bibinfo {author}
  {\bibfnamefont {S.}~\bibnamefont {Guillot}}, \bibinfo {author} {\bibfnamefont
  {Z.}~\bibnamefont {Arzoumanian}}, \bibinfo {author} {\bibfnamefont {C.~L.}\
  \bibnamefont {Baker}}, \bibinfo {author} {\bibfnamefont {A.~V.}\ \bibnamefont
  {Bilous}}, \bibinfo {author} {\bibfnamefont {D.}~\bibnamefont {Chakrabarty}},
  \bibinfo {author} {\bibfnamefont {K.~C.}\ \bibnamefont {Gendreau}}, \bibinfo
  {author} {\bibfnamefont {A.~K.}\ \bibnamefont {Harding}}, \bibinfo {author}
  {\bibfnamefont {W.~C.~G.}\ \bibnamefont {Ho}}, \bibinfo {author}
  {\bibfnamefont {J.~M.}\ \bibnamefont {Lattimer}}, \bibinfo {author}
  {\bibfnamefont {S.~M.}\ \bibnamefont {Morsink}}, \ and\ \bibinfo {author}
  {\bibfnamefont {T.~E.}\ \bibnamefont {Strohmayer}},\ }\href {\doibase
  10.3847/2041-8213/ab481c} {\bibfield  {journal} {\bibinfo  {journal} {The
  Astrophysical Journal}\ }\textbf {\bibinfo {volume} {887}},\ \bibinfo {pages}
  {L21} (\bibinfo {year} {2019})}\BibitemShut {NoStop}%
\bibitem [{\citenamefont {Capano}\ \emph {et~al.}(2020)\citenamefont {Capano},
  \citenamefont {Tews}, \citenamefont {Brown}, \citenamefont {Margalit},
  \citenamefont {De}, \citenamefont {Kumar}, \citenamefont {Brown},
  \citenamefont {Krishnan},\ and\ \citenamefont {Reddy}}]{Capano_2020}%
  \BibitemOpen
  \bibfield  {author} {\bibinfo {author} {\bibfnamefont {C.~D.}\ \bibnamefont
  {Capano}}, \bibinfo {author} {\bibfnamefont {I.}~\bibnamefont {Tews}},
  \bibinfo {author} {\bibfnamefont {S.~M.}\ \bibnamefont {Brown}}, \bibinfo
  {author} {\bibfnamefont {B.}~\bibnamefont {Margalit}}, \bibinfo {author}
  {\bibfnamefont {S.}~\bibnamefont {De}}, \bibinfo {author} {\bibfnamefont
  {S.}~\bibnamefont {Kumar}}, \bibinfo {author} {\bibfnamefont {D.~A.}\
  \bibnamefont {Brown}}, \bibinfo {author} {\bibfnamefont {B.}~\bibnamefont
  {Krishnan}}, \ and\ \bibinfo {author} {\bibfnamefont {S.}~\bibnamefont
  {Reddy}},\ }\href {\doibase 10.1038/s41550-020-1014-6} {\bibfield  {journal}
  {\bibinfo  {journal} {Nature Astronomy}\ }\textbf {\bibinfo {volume} {4}},\
  \bibinfo {pages} {625–632} (\bibinfo {year} {2020})}\BibitemShut {NoStop}%
\bibitem [{\citenamefont {Li}\ \emph {et~al.}(2018{\natexlab{a}})\citenamefont
  {Li}, \citenamefont {Yan}, \citenamefont {Geng}, \citenamefont {Huang},\ and\
  \citenamefont {Zong}}]{PhysRevD.98.083013}%
  \BibitemOpen
  \bibfield  {author} {\bibinfo {author} {\bibfnamefont {C.-M.}\ \bibnamefont
  {Li}}, \bibinfo {author} {\bibfnamefont {Y.}~\bibnamefont {Yan}}, \bibinfo
  {author} {\bibfnamefont {J.-J.}\ \bibnamefont {Geng}}, \bibinfo {author}
  {\bibfnamefont {Y.-F.}\ \bibnamefont {Huang}}, \ and\ \bibinfo {author}
  {\bibfnamefont {H.-S.}\ \bibnamefont {Zong}},\ }\href {\doibase
  10.1103/PhysRevD.98.083013} {\bibfield  {journal} {\bibinfo  {journal} {Phys.
  Rev. D}\ }\textbf {\bibinfo {volume} {98}},\ \bibinfo {pages} {083013}
  (\bibinfo {year} {2018}{\natexlab{a}})}\BibitemShut {NoStop}%
\bibitem [{\citenamefont {Li}\ \emph {et~al.}(2018{\natexlab{b}})\citenamefont
  {Li}, \citenamefont {Zhang}, \citenamefont {Yan}, \citenamefont {Huang},\
  and\ \citenamefont {Zong}}]{PhysRevD.97.103013}%
  \BibitemOpen
  \bibfield  {author} {\bibinfo {author} {\bibfnamefont {C.-M.}\ \bibnamefont
  {Li}}, \bibinfo {author} {\bibfnamefont {J.-L.}\ \bibnamefont {Zhang}},
  \bibinfo {author} {\bibfnamefont {Y.}~\bibnamefont {Yan}}, \bibinfo {author}
  {\bibfnamefont {Y.-F.}\ \bibnamefont {Huang}}, \ and\ \bibinfo {author}
  {\bibfnamefont {H.-S.}\ \bibnamefont {Zong}},\ }\href {\doibase
  10.1103/PhysRevD.97.103013} {\bibfield  {journal} {\bibinfo  {journal} {Phys.
  Rev. D}\ }\textbf {\bibinfo {volume} {97}},\ \bibinfo {pages} {103013}
  (\bibinfo {year} {2018}{\natexlab{b}})}\BibitemShut {NoStop}%
\bibitem [{\citenamefont {Zhao}\ \emph {et~al.}(2015)\citenamefont {Zhao},
  \citenamefont {Xu}, \citenamefont {Yan}, \citenamefont {Luo}, \citenamefont
  {Liu},\ and\ \citenamefont {Zong}}]{PhysRevD.92.054012}%
  \BibitemOpen
  \bibfield  {author} {\bibinfo {author} {\bibfnamefont {T.}~\bibnamefont
  {Zhao}}, \bibinfo {author} {\bibfnamefont {S.-S.}\ \bibnamefont {Xu}},
  \bibinfo {author} {\bibfnamefont {Y.}~\bibnamefont {Yan}}, \bibinfo {author}
  {\bibfnamefont {X.-L.}\ \bibnamefont {Luo}}, \bibinfo {author} {\bibfnamefont
  {X.-J.}\ \bibnamefont {Liu}}, \ and\ \bibinfo {author} {\bibfnamefont
  {H.-S.}\ \bibnamefont {Zong}},\ }\href {\doibase 10.1103/PhysRevD.92.054012}
  {\bibfield  {journal} {\bibinfo  {journal} {Phys. Rev. D}\ }\textbf {\bibinfo
  {volume} {92}},\ \bibinfo {pages} {054012} (\bibinfo {year}
  {2015})}\BibitemShut {NoStop}%
\bibitem [{\citenamefont {Witten}(1984)}]{PhysRevD.30.272}%
  \BibitemOpen
  \bibfield  {author} {\bibinfo {author} {\bibfnamefont {E.}~\bibnamefont
  {Witten}},\ }\href {\doibase 10.1103/PhysRevD.30.272} {\bibfield  {journal}
  {\bibinfo  {journal} {Phys. Rev. D}\ }\textbf {\bibinfo {volume} {30}},\
  \bibinfo {pages} {272} (\bibinfo {year} {1984})}\BibitemShut {NoStop}%
\bibitem [{\citenamefont {Menezes}\ \emph {et~al.}(2014)\citenamefont
  {Menezes}, \citenamefont {Pinto}, \citenamefont {Castro}, \citenamefont
  {Costa},\ and\ \citenamefont {Provid\^encia}}]{PhysRevC.89.055207}%
  \BibitemOpen
  \bibfield  {author} {\bibinfo {author} {\bibfnamefont {D.~P.}\ \bibnamefont
  {Menezes}}, \bibinfo {author} {\bibfnamefont {M.~B.}\ \bibnamefont {Pinto}},
  \bibinfo {author} {\bibfnamefont {L.~B.}\ \bibnamefont {Castro}}, \bibinfo
  {author} {\bibfnamefont {P.}~\bibnamefont {Costa}}, \ and\ \bibinfo {author}
  {\bibfnamefont {C.~m.~c.}\ \bibnamefont {Provid\^encia}},\ }\href {\doibase
  10.1103/PhysRevC.89.055207} {\bibfield  {journal} {\bibinfo  {journal} {Phys.
  Rev. C}\ }\textbf {\bibinfo {volume} {89}},\ \bibinfo {pages} {055207}
  (\bibinfo {year} {2014})}\BibitemShut {NoStop}%
\bibitem [{\citenamefont {Li}\ \emph {et~al.}(2020)\citenamefont {Li},
  \citenamefont {Zuo}, \citenamefont {Yan}, \citenamefont {Zhao}, \citenamefont
  {Wang}, \citenamefont {Huang},\ and\ \citenamefont
  {Zong}}]{PhysRevD.101.063023}%
  \BibitemOpen
  \bibfield  {author} {\bibinfo {author} {\bibfnamefont {C.-M.}\ \bibnamefont
  {Li}}, \bibinfo {author} {\bibfnamefont {S.-Y.}\ \bibnamefont {Zuo}},
  \bibinfo {author} {\bibfnamefont {Y.}~\bibnamefont {Yan}}, \bibinfo {author}
  {\bibfnamefont {Y.-P.}\ \bibnamefont {Zhao}}, \bibinfo {author}
  {\bibfnamefont {F.}~\bibnamefont {Wang}}, \bibinfo {author} {\bibfnamefont
  {Y.-F.}\ \bibnamefont {Huang}}, \ and\ \bibinfo {author} {\bibfnamefont
  {H.-S.}\ \bibnamefont {Zong}},\ }\href {\doibase 10.1103/PhysRevD.101.063023}
  {\bibfield  {journal} {\bibinfo  {journal} {Phys. Rev. D}\ }\textbf {\bibinfo
  {volume} {101}},\ \bibinfo {pages} {063023} (\bibinfo {year}
  {2020})}\BibitemShut {NoStop}%
\bibitem [{\citenamefont {Chakrabarty}(1991)}]{PhysRevD.43.627}%
  \BibitemOpen
  \bibfield  {author} {\bibinfo {author} {\bibfnamefont {S.}~\bibnamefont
  {Chakrabarty}},\ }\href {\doibase 10.1103/PhysRevD.43.627} {\bibfield
  {journal} {\bibinfo  {journal} {Phys. Rev. D}\ }\textbf {\bibinfo {volume}
  {43}},\ \bibinfo {pages} {627} (\bibinfo {year} {1991})}\BibitemShut
  {NoStop}%
\bibitem [{\citenamefont {Peng}\ \emph {et~al.}(2000)\citenamefont {Peng},
  \citenamefont {Chiang}, \citenamefont {Zou}, \citenamefont {Ning},\ and\
  \citenamefont {Luo}}]{PhysRevC.62.025801}%
  \BibitemOpen
  \bibfield  {author} {\bibinfo {author} {\bibfnamefont {G.~X.}\ \bibnamefont
  {Peng}}, \bibinfo {author} {\bibfnamefont {H.~C.}\ \bibnamefont {Chiang}},
  \bibinfo {author} {\bibfnamefont {B.~S.}\ \bibnamefont {Zou}}, \bibinfo
  {author} {\bibfnamefont {P.~Z.}\ \bibnamefont {Ning}}, \ and\ \bibinfo
  {author} {\bibfnamefont {S.~J.}\ \bibnamefont {Luo}},\ }\href {\doibase
  10.1103/PhysRevC.62.025801} {\bibfield  {journal} {\bibinfo  {journal} {Phys.
  Rev. C}\ }\textbf {\bibinfo {volume} {62}},\ \bibinfo {pages} {025801}
  (\bibinfo {year} {2000})}\BibitemShut {NoStop}%
\bibitem [{\citenamefont {Arba\~nil}\ and\ \citenamefont
  {Malheiro}(2015)}]{PhysRevD.92.084009}%
  \BibitemOpen
  \bibfield  {author} {\bibinfo {author} {\bibfnamefont {J.~D.~V.}\
  \bibnamefont {Arba\~nil}}\ and\ \bibinfo {author} {\bibfnamefont
  {M.}~\bibnamefont {Malheiro}},\ }\href {\doibase 10.1103/PhysRevD.92.084009}
  {\bibfield  {journal} {\bibinfo  {journal} {Phys. Rev. D}\ }\textbf {\bibinfo
  {volume} {92}},\ \bibinfo {pages} {084009} (\bibinfo {year}
  {2015})}\BibitemShut {NoStop}%
\bibitem [{\citenamefont {Li}\ \emph {et~al.}(2019)\citenamefont {Li},
  \citenamefont {Cui}, \citenamefont {Yu}, \citenamefont {Yan}, \citenamefont
  {An},\ and\ \citenamefont {Zong}}]{PhysRevD.99.043001}%
  \BibitemOpen
  \bibfield  {author} {\bibinfo {author} {\bibfnamefont {B.-L.}\ \bibnamefont
  {Li}}, \bibinfo {author} {\bibfnamefont {Z.-F.}\ \bibnamefont {Cui}},
  \bibinfo {author} {\bibfnamefont {Z.-H.}\ \bibnamefont {Yu}}, \bibinfo
  {author} {\bibfnamefont {Y.}~\bibnamefont {Yan}}, \bibinfo {author}
  {\bibfnamefont {S.}~\bibnamefont {An}}, \ and\ \bibinfo {author}
  {\bibfnamefont {H.-S.}\ \bibnamefont {Zong}},\ }\href {\doibase
  10.1103/PhysRevD.99.043001} {\bibfield  {journal} {\bibinfo  {journal} {Phys.
  Rev. D}\ }\textbf {\bibinfo {volume} {99}},\ \bibinfo {pages} {043001}
  (\bibinfo {year} {2019})}\BibitemShut {NoStop}%
\bibitem [{\citenamefont {Zhao}\ \emph {et~al.}(2019)\citenamefont {Zhao},
  \citenamefont {Zheng}, \citenamefont {Wang}, \citenamefont {Li},
  \citenamefont {Yan}, \citenamefont {Huang},\ and\ \citenamefont
  {Zong}}]{PhysRevD.100.043018}%
  \BibitemOpen
  \bibfield  {author} {\bibinfo {author} {\bibfnamefont {T.}~\bibnamefont
  {Zhao}}, \bibinfo {author} {\bibfnamefont {W.}~\bibnamefont {Zheng}},
  \bibinfo {author} {\bibfnamefont {F.}~\bibnamefont {Wang}}, \bibinfo {author}
  {\bibfnamefont {C.-M.}\ \bibnamefont {Li}}, \bibinfo {author} {\bibfnamefont
  {Y.}~\bibnamefont {Yan}}, \bibinfo {author} {\bibfnamefont {Y.-F.}\
  \bibnamefont {Huang}}, \ and\ \bibinfo {author} {\bibfnamefont {H.-S.}\
  \bibnamefont {Zong}},\ }\href {\doibase 10.1103/PhysRevD.100.043018}
  {\bibfield  {journal} {\bibinfo  {journal} {Phys. Rev. D}\ }\textbf {\bibinfo
  {volume} {100}},\ \bibinfo {pages} {043018} (\bibinfo {year}
  {2019})}\BibitemShut {NoStop}%
\bibitem [{\citenamefont {Wang}\ \emph
  {et~al.}(2019{\natexlab{a}})\citenamefont {Wang}, \citenamefont {Shi},\ and\
  \citenamefont {Zong}}]{PhysRevD.100.123003}%
  \BibitemOpen
  \bibfield  {author} {\bibinfo {author} {\bibfnamefont {Q.}~\bibnamefont
  {Wang}}, \bibinfo {author} {\bibfnamefont {C.}~\bibnamefont {Shi}}, \ and\
  \bibinfo {author} {\bibfnamefont {H.-S.}\ \bibnamefont {Zong}},\ }\href
  {\doibase 10.1103/PhysRevD.100.123003} {\bibfield  {journal} {\bibinfo
  {journal} {Phys. Rev. D}\ }\textbf {\bibinfo {volume} {100}},\ \bibinfo
  {pages} {123003} (\bibinfo {year} {2019}{\natexlab{a}})}\BibitemShut
  {NoStop}%
\bibitem [{\citenamefont {Holdom}\ \emph {et~al.}(2018)\citenamefont {Holdom},
  \citenamefont {Ren},\ and\ \citenamefont {Zhang}}]{PhysRevLett.120.222001}%
  \BibitemOpen
  \bibfield  {author} {\bibinfo {author} {\bibfnamefont {B.}~\bibnamefont
  {Holdom}}, \bibinfo {author} {\bibfnamefont {J.}~\bibnamefont {Ren}}, \ and\
  \bibinfo {author} {\bibfnamefont {C.}~\bibnamefont {Zhang}},\ }\href
  {\doibase 10.1103/PhysRevLett.120.222001} {\bibfield  {journal} {\bibinfo
  {journal} {Phys. Rev. Lett.}\ }\textbf {\bibinfo {volume} {120}},\ \bibinfo
  {pages} {222001} (\bibinfo {year} {2018})}\BibitemShut {NoStop}%
\bibitem [{\citenamefont {Zhu}\ \emph {et~al.}(2019)\citenamefont {Zhu},
  \citenamefont {Li}, \citenamefont {Hu},\ and\ \citenamefont
  {Shen}}]{PhysRevC.99.025804}%
  \BibitemOpen
  \bibfield  {author} {\bibinfo {author} {\bibfnamefont {Z.-Y.}\ \bibnamefont
  {Zhu}}, \bibinfo {author} {\bibfnamefont {A.}~\bibnamefont {Li}}, \bibinfo
  {author} {\bibfnamefont {J.-N.}\ \bibnamefont {Hu}}, \ and\ \bibinfo {author}
  {\bibfnamefont {H.}~\bibnamefont {Shen}},\ }\href {\doibase
  10.1103/PhysRevC.99.025804} {\bibfield  {journal} {\bibinfo  {journal} {Phys.
  Rev. C}\ }\textbf {\bibinfo {volume} {99}},\ \bibinfo {pages} {025804}
  (\bibinfo {year} {2019})}\BibitemShut {NoStop}%
\bibitem [{\citenamefont {Halasz}\ \emph {et~al.}(1998)\citenamefont {Halasz},
  \citenamefont {Jackson}, \citenamefont {Shrock}, \citenamefont {Stephanov},\
  and\ \citenamefont {Verbaarschot}}]{PhysRevD.58.096007}%
  \BibitemOpen
  \bibfield  {author} {\bibinfo {author} {\bibfnamefont {M.~A.}\ \bibnamefont
  {Halasz}}, \bibinfo {author} {\bibfnamefont {A.~D.}\ \bibnamefont {Jackson}},
  \bibinfo {author} {\bibfnamefont {R.~E.}\ \bibnamefont {Shrock}}, \bibinfo
  {author} {\bibfnamefont {M.~A.}\ \bibnamefont {Stephanov}}, \ and\ \bibinfo
  {author} {\bibfnamefont {J.~J.~M.}\ \bibnamefont {Verbaarschot}},\ }\href
  {\doibase 10.1103/PhysRevD.58.096007} {\bibfield  {journal} {\bibinfo
  {journal} {Phys. Rev. D}\ }\textbf {\bibinfo {volume} {58}},\ \bibinfo
  {pages} {096007} (\bibinfo {year} {1998})}\BibitemShut {NoStop}%
\bibitem [{\citenamefont {Kapusta}\ and\ \citenamefont
  {Gale}(2011)}]{Kapusta:2006pm}%
  \BibitemOpen
  \bibfield  {author} {\bibinfo {author} {\bibfnamefont {J.}~\bibnamefont
  {Kapusta}}\ and\ \bibinfo {author} {\bibfnamefont {C.}~\bibnamefont {Gale}},\
  }\href {\doibase 10.1017/CBO9780511535130} {\emph {\bibinfo {title}
  {{Finite-temperature field theory: Principles and applications}}}},\
  Cambridge Monographs on Mathematical Physics\ (\bibinfo  {publisher}
  {Cambridge University Press},\ \bibinfo {year} {2011})\BibitemShut {NoStop}%
\bibitem [{\citenamefont {Zong}\ and\ \citenamefont
  {Sun}(2008{\natexlab{a}})}]{PhysRevD.78.054001}%
  \BibitemOpen
  \bibfield  {author} {\bibinfo {author} {\bibfnamefont {H.-s.}\ \bibnamefont
  {Zong}}\ and\ \bibinfo {author} {\bibfnamefont {W.-m.}\ \bibnamefont {Sun}},\
  }\href {\doibase 10.1103/PhysRevD.78.054001} {\bibfield  {journal} {\bibinfo
  {journal} {Phys. Rev. D}\ }\textbf {\bibinfo {volume} {78}},\ \bibinfo
  {pages} {054001} (\bibinfo {year} {2008}{\natexlab{a}})}\BibitemShut
  {NoStop}%
\bibitem [{\citenamefont {Zong}\ and\ \citenamefont
  {Sun}(2008{\natexlab{b}})}]{Zong:2008zzb}%
  \BibitemOpen
  \bibfield  {author} {\bibinfo {author} {\bibfnamefont {H.-S.}\ \bibnamefont
  {Zong}}\ and\ \bibinfo {author} {\bibfnamefont {W.-M.}\ \bibnamefont {Sun}},\
  }\href {\doibase 10.1142/S0217751X08040457} {\bibfield  {journal} {\bibinfo
  {journal} {Int. J. Mod. Phys.}\ }\textbf {\bibinfo {volume} {A23}},\ \bibinfo
  {pages} {3591} (\bibinfo {year} {2008}{\natexlab{b}})}\BibitemShut {NoStop}%
\bibitem [{\citenamefont {Serot}\ and\ \citenamefont
  {Walecka}(1986{\natexlab{b}})}]{walecka1986}%
  \BibitemOpen
  \bibfield  {author} {\bibinfo {author} {\bibfnamefont {B.~D.}\ \bibnamefont
  {Serot}}\ and\ \bibinfo {author} {\bibfnamefont {J.~D.}\ \bibnamefont
  {Walecka}},\ }\href@noop {} {\emph {\bibinfo {title} {{Advances in Nuclear
  Physics}}}}\ (\bibinfo {year} {1986})\BibitemShut {NoStop}%
\bibitem [{\citenamefont {Fetter}\ and\ \citenamefont
  {Walecka}(2003)}]{fetter2003quantum}%
  \BibitemOpen
  \bibfield  {author} {\bibinfo {author} {\bibfnamefont {A.}~\bibnamefont
  {Fetter}}\ and\ \bibinfo {author} {\bibfnamefont {J.}~\bibnamefont
  {Walecka}},\ }\href {https://books.google.com/books?id=0wekf1s83b0C} {\emph
  {\bibinfo {title} {Quantum Theory of Many-particle Systems}}},\ Dover Books
  on Physics\ (\bibinfo  {publisher} {Dover Publications},\ \bibinfo {year}
  {2003})\BibitemShut {NoStop}%
\bibitem [{\citenamefont {Dutra}\ \emph {et~al.}(2014)\citenamefont {Dutra},
  \citenamefont {Louren\ifmmode~\mbox{\c{c}}\else \c{c}\fi{}o}, \citenamefont
  {Avancini}, \citenamefont {Carlson}, \citenamefont {Delfino}, \citenamefont
  {Menezes}, \citenamefont {Provid\^encia}, \citenamefont {Typel},\ and\
  \citenamefont {Stone}}]{PhysRevC.90.055203}%
  \BibitemOpen
  \bibfield  {author} {\bibinfo {author} {\bibfnamefont {M.}~\bibnamefont
  {Dutra}}, \bibinfo {author} {\bibfnamefont {O.}~\bibnamefont
  {Louren\ifmmode~\mbox{\c{c}}\else \c{c}\fi{}o}}, \bibinfo {author}
  {\bibfnamefont {S.~S.}\ \bibnamefont {Avancini}}, \bibinfo {author}
  {\bibfnamefont {B.~V.}\ \bibnamefont {Carlson}}, \bibinfo {author}
  {\bibfnamefont {A.}~\bibnamefont {Delfino}}, \bibinfo {author} {\bibfnamefont
  {D.~P.}\ \bibnamefont {Menezes}}, \bibinfo {author} {\bibfnamefont
  {C.}~\bibnamefont {Provid\^encia}}, \bibinfo {author} {\bibfnamefont
  {S.}~\bibnamefont {Typel}}, \ and\ \bibinfo {author} {\bibfnamefont {J.~R.}\
  \bibnamefont {Stone}},\ }\href {\doibase 10.1103/PhysRevC.90.055203}
  {\bibfield  {journal} {\bibinfo  {journal} {Phys. Rev. C}\ }\textbf {\bibinfo
  {volume} {90}},\ \bibinfo {pages} {055203} (\bibinfo {year}
  {2014})}\BibitemShut {NoStop}%
\bibitem [{\citenamefont {{Fetter}}\ \emph {et~al.}(1972)\citenamefont
  {{Fetter}}, \citenamefont {{Walecka}},\ and\ \citenamefont
  {{Kadanoff}}}]{1972PhT....25k..54F}%
  \BibitemOpen
  \bibfield  {author} {\bibinfo {author} {\bibfnamefont {A.~L.}\ \bibnamefont
  {{Fetter}}}, \bibinfo {author} {\bibfnamefont {J.~D.}\ \bibnamefont
  {{Walecka}}}, \ and\ \bibinfo {author} {\bibfnamefont {L.~P.}\ \bibnamefont
  {{Kadanoff}}},\ }\href {\doibase 10.1063/1.3071096} {\bibfield  {journal}
  {\bibinfo  {journal} {Physics Today}\ }\textbf {\bibinfo {volume} {25}},\
  \bibinfo {pages} {54} (\bibinfo {year} {1972})}\BibitemShut {NoStop}%
\bibitem [{\citenamefont {Li}\ \emph {et~al.}(2008)\citenamefont {Li},
  \citenamefont {Chen},\ and\ \citenamefont {Ko}}]{LI2008113}%
  \BibitemOpen
  \bibfield  {author} {\bibinfo {author} {\bibfnamefont {B.-A.}\ \bibnamefont
  {Li}}, \bibinfo {author} {\bibfnamefont {L.-W.}\ \bibnamefont {Chen}}, \ and\
  \bibinfo {author} {\bibfnamefont {C.~M.}\ \bibnamefont {Ko}},\ }\href
  {\doibase https://doi.org/10.1016/j.physrep.2008.04.005} {\bibfield
  {journal} {\bibinfo  {journal} {Physics Reports}\ }\textbf {\bibinfo {volume}
  {464}},\ \bibinfo {pages} {113 } (\bibinfo {year} {2008})}\BibitemShut
  {NoStop}%
\bibitem [{\citenamefont {Fukushima}\ and\ \citenamefont
  {Sasaki}(2013)}]{FUKUSHIMA201399}%
  \BibitemOpen
  \bibfield  {author} {\bibinfo {author} {\bibfnamefont {K.}~\bibnamefont
  {Fukushima}}\ and\ \bibinfo {author} {\bibfnamefont {C.}~\bibnamefont
  {Sasaki}},\ }\href {\doibase https://doi.org/10.1016/j.ppnp.2013.05.003}
  {\bibfield  {journal} {\bibinfo  {journal} {Progress in Particle and Nuclear
  Physics}\ }\textbf {\bibinfo {volume} {72}},\ \bibinfo {pages} {99 }
  (\bibinfo {year} {2013})}\BibitemShut {NoStop}%
\bibitem [{\citenamefont {Chodos}\ \emph {et~al.}(1974)\citenamefont {Chodos},
  \citenamefont {Jaffe}, \citenamefont {Johnson}, \citenamefont {Thorn},\ and\
  \citenamefont {Weisskopf}}]{PhysRevD.9.3471}%
  \BibitemOpen
  \bibfield  {author} {\bibinfo {author} {\bibfnamefont {A.}~\bibnamefont
  {Chodos}}, \bibinfo {author} {\bibfnamefont {R.~L.}\ \bibnamefont {Jaffe}},
  \bibinfo {author} {\bibfnamefont {K.}~\bibnamefont {Johnson}}, \bibinfo
  {author} {\bibfnamefont {C.~B.}\ \bibnamefont {Thorn}}, \ and\ \bibinfo
  {author} {\bibfnamefont {V.~F.}\ \bibnamefont {Weisskopf}},\ }\href {\doibase
  10.1103/PhysRevD.9.3471} {\bibfield  {journal} {\bibinfo  {journal} {Phys.
  Rev. D}\ }\textbf {\bibinfo {volume} {9}},\ \bibinfo {pages} {3471} (\bibinfo
  {year} {1974})}\BibitemShut {NoStop}%
\bibitem [{\citenamefont {Zhao}\ \emph {et~al.}(2014)\citenamefont {Zhao},
  \citenamefont {Cui}, \citenamefont {Jiang},\ and\ \citenamefont
  {Zong}}]{PhysRevD.90.114031}%
  \BibitemOpen
  \bibfield  {author} {\bibinfo {author} {\bibfnamefont {A.-M.}\ \bibnamefont
  {Zhao}}, \bibinfo {author} {\bibfnamefont {Z.-F.}\ \bibnamefont {Cui}},
  \bibinfo {author} {\bibfnamefont {Y.}~\bibnamefont {Jiang}}, \ and\ \bibinfo
  {author} {\bibfnamefont {H.-S.}\ \bibnamefont {Zong}},\ }\href {\doibase
  10.1103/PhysRevD.90.114031} {\bibfield  {journal} {\bibinfo  {journal} {Phys.
  Rev. D}\ }\textbf {\bibinfo {volume} {90}},\ \bibinfo {pages} {114031}
  (\bibinfo {year} {2014})}\BibitemShut {NoStop}%
\bibitem [{\citenamefont {Wang}\ \emph {et~al.}(2015)\citenamefont {Wang},
  \citenamefont {Wang}, \citenamefont {Cui},\ and\ \citenamefont
  {Zong}}]{PhysRevD.91.034017}%
  \BibitemOpen
  \bibfield  {author} {\bibinfo {author} {\bibfnamefont {B.}~\bibnamefont
  {Wang}}, \bibinfo {author} {\bibfnamefont {Y.-L.}\ \bibnamefont {Wang}},
  \bibinfo {author} {\bibfnamefont {Z.-F.}\ \bibnamefont {Cui}}, \ and\
  \bibinfo {author} {\bibfnamefont {H.-S.}\ \bibnamefont {Zong}},\ }\href
  {\doibase 10.1103/PhysRevD.91.034017} {\bibfield  {journal} {\bibinfo
  {journal} {Phys. Rev. D}\ }\textbf {\bibinfo {volume} {91}},\ \bibinfo
  {pages} {034017} (\bibinfo {year} {2015})}\BibitemShut {NoStop}%
\bibitem [{\citenamefont {Xu}\ \emph {et~al.}(2015)\citenamefont {Xu},
  \citenamefont {Cui}, \citenamefont {Wang}, \citenamefont {Shi}, \citenamefont
  {Yang},\ and\ \citenamefont {Zong}}]{PhysRevD.91.056003}%
  \BibitemOpen
  \bibfield  {author} {\bibinfo {author} {\bibfnamefont {S.-S.}\ \bibnamefont
  {Xu}}, \bibinfo {author} {\bibfnamefont {Z.-F.}\ \bibnamefont {Cui}},
  \bibinfo {author} {\bibfnamefont {B.}~\bibnamefont {Wang}}, \bibinfo {author}
  {\bibfnamefont {Y.-M.}\ \bibnamefont {Shi}}, \bibinfo {author} {\bibfnamefont
  {Y.-C.}\ \bibnamefont {Yang}}, \ and\ \bibinfo {author} {\bibfnamefont
  {H.-S.}\ \bibnamefont {Zong}},\ }\href {\doibase 10.1103/PhysRevD.91.056003}
  {\bibfield  {journal} {\bibinfo  {journal} {Phys. Rev. D}\ }\textbf {\bibinfo
  {volume} {91}},\ \bibinfo {pages} {056003} (\bibinfo {year}
  {2015})}\BibitemShut {NoStop}%
\bibitem [{\citenamefont {Klevansky}(1992)}]{RevModPhys.64.649}%
  \BibitemOpen
  \bibfield  {author} {\bibinfo {author} {\bibfnamefont {S.~P.}\ \bibnamefont
  {Klevansky}},\ }\href {\doibase 10.1103/RevModPhys.64.649} {\bibfield
  {journal} {\bibinfo  {journal} {Rev. Mod. Phys.}\ }\textbf {\bibinfo {volume}
  {64}},\ \bibinfo {pages} {649} (\bibinfo {year} {1992})}\BibitemShut
  {NoStop}%
\bibitem [{\citenamefont {Masayuki}\ and\ \citenamefont
  {Koichi}(1989)}]{MASAYUKI1989668}%
  \BibitemOpen
  \bibfield  {author} {\bibinfo {author} {\bibfnamefont {A.}~\bibnamefont
  {Masayuki}}\ and\ \bibinfo {author} {\bibfnamefont {Y.}~\bibnamefont
  {Koichi}},\ }\href {\doibase https://doi.org/10.1016/0375-9474(89)90002-X}
  {\bibfield  {journal} {\bibinfo  {journal} {Nuclear Physics A}\ }\textbf
  {\bibinfo {volume} {504}},\ \bibinfo {pages} {668 } (\bibinfo {year}
  {1989})}\BibitemShut {NoStop}%
\bibitem [{\citenamefont {Cloët}\ and\ \citenamefont
  {Roberts}(2014)}]{CLOET20141}%
  \BibitemOpen
  \bibfield  {author} {\bibinfo {author} {\bibfnamefont {I.~C.}\ \bibnamefont
  {Cloët}}\ and\ \bibinfo {author} {\bibfnamefont {C.~D.}\ \bibnamefont
  {Roberts}},\ }\href {\doibase https://doi.org/10.1016/j.ppnp.2014.02.001}
  {\bibfield  {journal} {\bibinfo  {journal} {Progress in Particle and Nuclear
  Physics}\ }\textbf {\bibinfo {volume} {77}},\ \bibinfo {pages} {1 } (\bibinfo
  {year} {2014})}\BibitemShut {NoStop}%
\bibitem [{\citenamefont {Roberts}\ and\ \citenamefont
  {Schmidt}(2000)}]{ROBERTS2000S1}%
  \BibitemOpen
  \bibfield  {author} {\bibinfo {author} {\bibfnamefont {C.}~\bibnamefont
  {Roberts}}\ and\ \bibinfo {author} {\bibfnamefont {S.}~\bibnamefont
  {Schmidt}},\ }\href {\doibase https://doi.org/10.1016/S0146-6410(00)90011-5}
  {\bibfield  {journal} {\bibinfo  {journal} {Progress in Particle and Nuclear
  Physics}\ }\textbf {\bibinfo {volume} {45}},\ \bibinfo {pages} {S1 }
  (\bibinfo {year} {2000})}\BibitemShut {NoStop}%
\bibitem [{\citenamefont {Roberts}\ and\ \citenamefont
  {Williams}(1994)}]{ROBERTS1994477}%
  \BibitemOpen
  \bibfield  {author} {\bibinfo {author} {\bibfnamefont {C.~D.}\ \bibnamefont
  {Roberts}}\ and\ \bibinfo {author} {\bibfnamefont {A.~G.}\ \bibnamefont
  {Williams}},\ }\href {\doibase https://doi.org/10.1016/0146-6410(94)90049-3}
  {\bibfield  {journal} {\bibinfo  {journal} {Progress in Particle and Nuclear
  Physics}\ }\textbf {\bibinfo {volume} {33}},\ \bibinfo {pages} {477 }
  (\bibinfo {year} {1994})}\BibitemShut {NoStop}%
\bibitem [{\citenamefont {Buballa}(2005)}]{Buballa:2003qv}%
  \BibitemOpen
  \bibfield  {author} {\bibinfo {author} {\bibfnamefont {M.}~\bibnamefont
  {Buballa}},\ }\href {\doibase 10.1016/j.physrep.2004.11.004} {\bibfield
  {journal} {\bibinfo  {journal} {Phys. Rept.}\ }\textbf {\bibinfo {volume}
  {407}},\ \bibinfo {pages} {205} (\bibinfo {year} {2005})}\BibitemShut
  {NoStop}%
\bibitem [{\citenamefont {Yu}\ \emph {et~al.}(2020)\citenamefont {Yu},
  \citenamefont {Zhao},\ and\ \citenamefont {Zong}}]{Yu:2020dnj}%
  \BibitemOpen
  \bibfield  {author} {\bibinfo {author} {\bibfnamefont {Z.-X.}\ \bibnamefont
  {Yu}}, \bibinfo {author} {\bibfnamefont {T.}~\bibnamefont {Zhao}}, \ and\
  \bibinfo {author} {\bibfnamefont {H.-S.}\ \bibnamefont {Zong}},\ }\href
  {\doibase 10.1088/1674-1137/44/7/074104} {\bibfield  {journal} {\bibinfo
  {journal} {Chin. Phys. C}\ }\textbf {\bibinfo {volume} {44}},\ \bibinfo
  {pages} {074104} (\bibinfo {year} {2020})}\BibitemShut {NoStop}%
\bibitem [{\citenamefont {Yang}\ \emph {et~al.}(2019)\citenamefont {Yang},
  \citenamefont {Luo},\ and\ \citenamefont {Zong}}]{PhysRevD.100.094012}%
  \BibitemOpen
  \bibfield  {author} {\bibinfo {author} {\bibfnamefont {L.-K.}\ \bibnamefont
  {Yang}}, \bibinfo {author} {\bibfnamefont {X.}~\bibnamefont {Luo}}, \ and\
  \bibinfo {author} {\bibfnamefont {H.-S.}\ \bibnamefont {Zong}},\ }\href
  {\doibase 10.1103/PhysRevD.100.094012} {\bibfield  {journal} {\bibinfo
  {journal} {Phys. Rev. D}\ }\textbf {\bibinfo {volume} {100}},\ \bibinfo
  {pages} {094012} (\bibinfo {year} {2019})}\BibitemShut {NoStop}%
\bibitem [{\citenamefont {Wang}\ \emph
  {et~al.}(2019{\natexlab{b}})\citenamefont {Wang}, \citenamefont {Cao},\ and\
  \citenamefont {Zong}}]{Wang_2019}%
  \BibitemOpen
  \bibfield  {author} {\bibinfo {author} {\bibfnamefont {F.}~\bibnamefont
  {Wang}}, \bibinfo {author} {\bibfnamefont {Y.}~\bibnamefont {Cao}}, \ and\
  \bibinfo {author} {\bibfnamefont {H.}~\bibnamefont {Zong}},\ }\href {\doibase
  10.1088/1674-1137/43/8/084102} {\bibfield  {journal} {\bibinfo  {journal}
  {Chinese Physics C}\ }\textbf {\bibinfo {volume} {43}},\ \bibinfo {pages}
  {084102} (\bibinfo {year} {2019}{\natexlab{b}})}\BibitemShut {NoStop}%
\bibitem [{\citenamefont {Su}\ \emph {et~al.}(2020)\citenamefont {Su},
  \citenamefont {Shi}, \citenamefont {Xia},\ and\ \citenamefont
  {Zong}}]{PhysRevD.102.054028}%
  \BibitemOpen
  \bibfield  {author} {\bibinfo {author} {\bibfnamefont {L.-Q.}\ \bibnamefont
  {Su}}, \bibinfo {author} {\bibfnamefont {C.}~\bibnamefont {Shi}}, \bibinfo
  {author} {\bibfnamefont {Y.-H.}\ \bibnamefont {Xia}}, \ and\ \bibinfo
  {author} {\bibfnamefont {H.}~\bibnamefont {Zong}},\ }\href {\doibase
  10.1103/PhysRevD.102.054028} {\bibfield  {journal} {\bibinfo  {journal}
  {Phys. Rev. D}\ }\textbf {\bibinfo {volume} {102}},\ \bibinfo {pages}
  {054028} (\bibinfo {year} {2020})}\BibitemShut {NoStop}%
\bibitem [{\citenamefont {Whittenbury}\ \emph {et~al.}(2016)\citenamefont
  {Whittenbury}, \citenamefont {Matevosyan},\ and\ \citenamefont
  {Thomas}}]{PhysRevC.93.035807}%
  \BibitemOpen
  \bibfield  {author} {\bibinfo {author} {\bibfnamefont {D.~L.}\ \bibnamefont
  {Whittenbury}}, \bibinfo {author} {\bibfnamefont {H.~H.}\ \bibnamefont
  {Matevosyan}}, \ and\ \bibinfo {author} {\bibfnamefont {A.~W.}\ \bibnamefont
  {Thomas}},\ }\href {\doibase 10.1103/PhysRevC.93.035807} {\bibfield
  {journal} {\bibinfo  {journal} {Phys. Rev. C}\ }\textbf {\bibinfo {volume}
  {93}},\ \bibinfo {pages} {035807} (\bibinfo {year} {2016})}\BibitemShut
  {NoStop}%
\bibitem [{\citenamefont {Kojo}\ \emph {et~al.}(2016)\citenamefont {Kojo},
  \citenamefont {Powell}, \citenamefont {Song},\ and\ \citenamefont
  {Baym}}]{KOJO2016821}%
  \BibitemOpen
  \bibfield  {author} {\bibinfo {author} {\bibfnamefont {T.}~\bibnamefont
  {Kojo}}, \bibinfo {author} {\bibfnamefont {P.~D.}\ \bibnamefont {Powell}},
  \bibinfo {author} {\bibfnamefont {Y.}~\bibnamefont {Song}}, \ and\ \bibinfo
  {author} {\bibfnamefont {G.}~\bibnamefont {Baym}},\ }\href {\doibase
  https://doi.org/10.1016/j.nuclphysa.2016.02.008} {\bibfield  {journal}
  {\bibinfo  {journal} {Nuclear Physics A}\ }\textbf {\bibinfo {volume}
  {956}},\ \bibinfo {pages} {821 } (\bibinfo {year} {2016})},\ \bibinfo {note}
  {the XXV International Conference on Ultrarelativistic Nucleus-Nucleus
  Collisions: Quark Matter 2015}\BibitemShut {NoStop}%
\bibitem [{\citenamefont {Masuda}\ \emph {et~al.}(2013)\citenamefont {Masuda},
  \citenamefont {Hatsuda},\ and\ \citenamefont
  {Takatsuka}}]{10.1093/ptep/ptt045}%
  \BibitemOpen
  \bibfield  {author} {\bibinfo {author} {\bibfnamefont {K.}~\bibnamefont
  {Masuda}}, \bibinfo {author} {\bibfnamefont {T.}~\bibnamefont {Hatsuda}}, \
  and\ \bibinfo {author} {\bibfnamefont {T.}~\bibnamefont {Takatsuka}},\ }\href
  {\doibase 10.1093/ptep/ptt045} {\bibfield  {journal} {\bibinfo  {journal}
  {Progress of Theoretical and Experimental Physics}\ }\textbf {\bibinfo
  {volume} {2013}} (\bibinfo {year} {2013}),\ 10.1093/ptep/ptt045},\ \bibinfo
  {note} {073D01},\ \Eprint
  {http://arxiv.org/abs/https://academic.oup.com/ptep/article-pdf/2013/7/073D01/19300294/ptt045.pdf}
  {https://academic.oup.com/ptep/article-pdf/2013/7/073D01/19300294/ptt045.pdf}
  \BibitemShut {NoStop}%
\bibitem [{\citenamefont {Hinderer}\ \emph {et~al.}(2010)\citenamefont
  {Hinderer}, \citenamefont {Lackey}, \citenamefont {Lang},\ and\ \citenamefont
  {Read}}]{PhysRevD.81.123016}%
  \BibitemOpen
  \bibfield  {author} {\bibinfo {author} {\bibfnamefont {T.}~\bibnamefont
  {Hinderer}}, \bibinfo {author} {\bibfnamefont {B.~D.}\ \bibnamefont
  {Lackey}}, \bibinfo {author} {\bibfnamefont {R.~N.}\ \bibnamefont {Lang}}, \
  and\ \bibinfo {author} {\bibfnamefont {J.~S.}\ \bibnamefont {Read}},\ }\href
  {\doibase 10.1103/PhysRevD.81.123016} {\bibfield  {journal} {\bibinfo
  {journal} {Phys. Rev. D}\ }\textbf {\bibinfo {volume} {81}},\ \bibinfo
  {pages} {123016} (\bibinfo {year} {2010})}\BibitemShut {NoStop}%
\end{thebibliography}%
\end{document}